\documentclass[aps,preprint,showpacs,showkeys,nofootinbib,floatfix,tightenlines,
superscriptaddress]{revtex4}

\usepackage{graphicx}  %
\usepackage{amsmath}
\usepackage{bm}  %

\newcommand{\bea}{\begin{eqnarray}}
\newcommand{\eea}{\end{eqnarray}}
\newcommand{\beq}{\begin{equation}}
\newcommand{\eeq}{\end{equation}}
\newcommand{\bqa}{\begin{eqnarray}}
\newcommand{\eqa}{\end{eqnarray}}

\newcommand{\bmAD}{A\bm{D}}
\def\mqo2{{\!\!\!}}

\begin{document}

\title{
Three-Body Recombination of Identical Bosons\\ 
with a Large Positive Scattering Length\\
at Nonzero Temperature}

\author{Eric Braaten}\email{braaten@mps.ohio-state.edu}
\affiliation{Department of Physics,
         The Ohio State University, Columbus, OH\ 43210, USA\\}

\author{H.-W. Hammer}\email{hammer@itkp.uni-bonn.de}
\affiliation{Helmholtz-Institut f\"ur Strahlen- und Kernphysik (Theorie),
 Universit\"at Bonn, 53115 Bonn, Germany\\}

\author{Daekyoung Kang}\email{kang@mps.ohio-state.edu}
\affiliation{Department of Physics,
         The Ohio State University, Columbus, OH\ 43210, USA\\}

\author{Lucas Platter}\email{platter.2@osu.edu}
\affiliation{Department of Physics,
         The Ohio State University, Columbus, OH\ 43210, USA\\}
\affiliation{Department of Physics and Astronomy, Ohio University,
        Athens, OH\ 45701, USA\\}
\date{\today}
%\date{November 2007}

\begin{abstract}
For identical bosons with a large scattering length,
the dependence of the 3-body recombination rate on the collision energy 
is determined in the zero-range limit 
by universal functions of a single scaling variable.
There are six scaling functions for angular momentum zero 
and one scaling function for each higher partial wave.
We calculate these universal functions by solving 
the Skorniakov--Ter-Martirosian equation.
The results for the 3-body recombination 
as a function of the collision energy 
are in good agreement with previous results from solving 
the 3-body Schr\"odinger equation for $^4$He atoms.
The universal scaling functions can be used to calculate the
3-body recombination rate at nonzero temperature.
We obtain an excellent fit to the data from the Innsbruck group for
$^{133}$Cs atoms with a large positive scattering length.
\end{abstract}

\smallskip
\maketitle

\section{Introduction}

Three-body recombination is a 3-particle process in which 
two of the particles bind to form a molecule.
This process is important in cold atom physics, because it is 
one of the most important loss mechanisms for trapped ultracold atoms.
If the interactions of the atoms are known with sufficient accuracy, 
the 3-body recombination rate can be calculated by solving 
the 3-body Schr\"odinger equation numerically.
However the interactions between complex atoms may not be known 
with sufficient accuracy.  This is especially true for  
atoms whose S-wave scattering length $a$ is large compared to 
the range of their interaction.  
In this case, the atoms have {\it universal} properties 
that depend on $a$ but are otherwise insensitive to length scales 
set by the range.  This makes possible a complementary approach 
to the few-body problem that involves expanding in powers of the 
range divided by $a$.   The leading term in this expansion is
called the {\it zero-range limit} or, alternatively, the 
{\it scaling limit}.  
In the 3-body sector, the approach to the 
few-body problem based on expanding around the zero-range limit
was pioneered by Vitaly Efimov beginning around 1970.

Efimov discovered that the zero-range limit of the 3-body problem for 
nonrelativistic particles with short-range interactions 
is quite remarkable.
If $a = \pm \infty$, there are infinitely many 3-body 
bound states with an accumulation point at the 3-atom
scattering threshold. These {\it Efimov states} or 
{\it Efimov trimers} have a geometric spectrum \cite{Efimov70}: 
%----------------------
\begin{eqnarray}
E^{(n)}_T = (e^{-2\pi/s_0})^{n-n_*} \hbar^2 \kappa^2_* /m,
\label{kappa-star}
\end{eqnarray}
%----------------------
%[hw,e,l,d] 
where $\kappa_*$ is the binding wavenumber of the  
Efimov trimer labeled by $n_*$ and $m$ is the mass of the
particles. This geometric spectrum 
is a signature of a {\it  discrete scaling symmetry}
with discrete scaling factor $e^{\pi/s_0}$ 
\cite{Braaten:2004rn}.
In the case of identical bosons, $s_0 \approx 1.00624$
and the discrete scaling factor is $e^{\pi/s_0} \approx 22.7$.
We will refer to the universal phenomena 
characterized by this discrete scaling symmetry as 
{\it Efimov physics} \cite{Braaten:2006vd}. 
Efimov showed that discrete scale invariance
is also relevant if $a$ is large but finite \cite{Efimov71,Efimov79}.
The negative values of $a$ for which 
there is an Efimov trimer at the 3-atom scattering threshold
are related by the discrete scaling symmetry.
The positive values of $a$ for which 
there is an Efimov trimer at the atom-dimer scattering threshold
are also related by the discrete scaling symmetry. 
Another dramatic example of Efimov physics is the vanishing of
the 3-body recombination rate at threshold
at positive values of $a$ that are related by the 
discrete scaling symmetry \cite{NM-99,EGB-99,Bedaque:2000ft}.

Efimov also showed that the discrete scaling symmetry governs the 
energy dependence of 3-body scattering processes, such as 
atom-dimer elastic scattering and 3-body recombination \cite{Efimov79}.
It is convenient to measure the total energy $E$ of the 
three atoms in their center-of-mass frame relative 
to the 3-atom scattering threshold.
The discrete scaling symmetry implies that the scattering rates 
at energy $E$ in a system with parameters $a$ and $\kappa_*$ 
differ from those at energy $(e^{-2 \pi/s_0})^n E$ in a system with 
parameters $(e^{\pi/s_0})^na$ and $\kappa_*$ 
only by an overall change in the scale.
The zero-range limit gives more detailed predictions
that Efimov referred to as the {\it Radial Law}.  The Radial Law expresses 
the dependence of the scattering rates on $E$ in terms of universal
functions of a scaling variable $x = (a^2 m E/\hbar^2)^{1/2}$.
Once these scaling functions have been calculated, 
the scattering rates can be predicted for any system of identical bosons 
with a large scattering length.  

The universal results defined by the scaling limit are exact 
only in the limit of zero range. 
Corrections to the universal results associated with a nonzero range
are suppressed by powers of $\ell/|a|$,
where $\ell$ is the natural low-energy length scale \cite{Braaten:2004rn}.
For a short-range potential, $\ell$ is simply the range 
of the potential.  For atoms whose potential has a van der Waals tail 
$-C_6/r^6$, $\ell$ is the van der Waals length
$\ell_{\rm vdW}=(mC_6/\hbar^2)^{1/4}$.
The effects of the nonzero range can be calculated
within a systematic expansion in the small parameter $\ell/|a|$ 
\cite{Efimov91,Efimov93,Hammer:2001gh,Bedaque:2002yg,Platter:2006ev}.
The next-to-leading order term in this expansion is proportional to 
$r_s/a$, where $r_s$ is the effective range \cite{Efimov91,Efimov93}.
Surprisingly, the next-to-next-to-leading order term in this expansion 
is also determined only by the scattering length $a$, 
the Efimov parameter $\kappa_*$, 
and the effective range \cite{Platter:2006ev}.

The natural low-energy length scale $\ell$ allows us to differentiate
between a {\it shallow dimer} and {\it deep dimers}. 
The shallow dimer
is a universal feature of the two-body system with large positive
scattering length and  has binding energy $E_D\approx \hbar^2/(m a^2)$. 
The underlying interaction can also support deep dimers whose
binding energies are comparable to or larger than $\hbar^2/(m \ell^2)$. 
The properties of these deep dimers are not universal.
The alkali atoms 
used in most cold atom experiments form diatomic molecules
with many deeply-bound energy levels.
Efimov trimers can therefore decay into an atom 
and a deep dimer with large kinetic energies.  
These final states also provide inelastic atom-dimer scattering 
channels and additional 3-body recombination channels.
If the atoms do not form deep dimers,
the zero-range results for scattering rates are determined by two parameters:
the scattering length $a$ and the Efimov parameter $\kappa_*$.
If the atoms do form deep dimers, the zero-range results for
scattering rates are determined by three parameters:
$a$, $\kappa_*$, and a parameter $\eta_*$ that determines 
the widths of the Efimov trimers \cite{Braaten:2002sr,Braaten:2003yc}.
Remarkably, the same universal functions that determine the energy 
dependence of scattering rates for the case $\eta_* = 0$
in which there are no deep dimers also determine their energy 
dependence for $\eta_* > 0$. 

The first experimental evidence for Efimov physics 
was presented by the Innsbruck group \cite{Grimm06}.
They carried out experiments with ultracold 
$^{133}$Cs atoms in the lowest hyperfine state,
using a magnetic field to control the scattering length. 
They observed a resonant enhancement in the 3-body recombination rate 
at $a \approx -850~a_0$ that can be attributed to an Efimov trimer 
near the 3-atom threshold.  At the temperature 10 nK, 
the loss rate as a function of $a$ can be fitted 
rather well by the universal formula for zero temperature 
derived in Ref.~\cite{Braaten:2003yc}
with a width parameter $\eta_* = 0.06(1)$.
The Innsbruck group also observed a local minimum in the 
3-body recombination rate near $a \approx 210~a_0$ that might be
associated with an interference minimum of the 3-body recombination rate. 
The measurements were carried out at 200 nK, which is large enough 
that taking into account the nonzero temperature is essential.

The 3-body recombination rate can be calculated at nonzero temperature 
by carrying out a thermal average of the 3-body recombination rate
as a function of the collision energy. 
There have been several previous calculations of the 3-body 
recombination rate for atoms with large scattering length 
at nonzero temperature using specific models.
D'Incao, Suno, and Esry used a simple 2-parameter potential with
either one or two S-wave bound states \cite{DSE-04,DSE-07}.
Massignan and Stoof used a 4-parameter scattering model for 
atoms near a Feshbach resonance \cite{MS07}.
There have been several previous calculations of the 3-body 
recombination rate for nonzero temperature that have exploited the 
universality of atoms with large scattering length.
In the case $a<0$, the 3-body recombination rate was 
calculated by Jonsell using the adiabatic hyperspherical approximation
\cite{Jonsell06} and by Yamashita, Frederico, and Tomio 
using a resonance approximation \cite{YFT06}.
In the case $a>0$, the 3-body recombination rate has been 
calculated using simplifying assumptions to neglect some of the 
universal scaling functions 
\cite{Braaten:2006qx,Shepard:2007gj,Platter:2007sn}.
The calculations in 
Refs.~\cite{Jonsell06,YFT06,Braaten:2006qx,Shepard:2007gj,Platter:2007sn} 
all involve uncontrolled approximations, so they are not definitive 
zero-range predictions.\footnote{Definitive zero-range predictions at
temperatures small compared to the dimer binding energy are
available for resonant dimer relaxation,
which is an important loss process for atom-dimer mixtures
\cite{Braaten:2006nn}.}

In this paper, we present definitive zero-range predictions  
for the 3-body recombination rate at nonzero temperature
for the case $a>0$.
We calculate the universal scaling functions 
that determine the hyperangular average of the
3-body recombination rate at collision energies 
up to about 30 times the binding energy of the shallow dimer.  
Our results are validated by comparing with a previous calculation
of the 3-body recombination rate as a function of collision energy
for $^4$He atoms.  
The universal scaling functions can be used to calculate 
the rate constants for 3-body recombination into the shallow dimer 
and into deep dimers at temperatures and densities where Boltzmann 
statistics are applicable.  We apply the results to 
$^{133}$Cs atoms with large positive scattering length $a$
and show that they give an excellent fit to the data from the 
Innsbruck group on the 3-body 
recombination rate as a function of $a$.

\section{Three-body Recombination}

Three-body recombination is a 3-atom collision process in which 
two of the atoms bind to form a diatomic molecule or {\it dimer}.
We take the 3 atoms to have the same mass $m$.
The 3-body recombination rate $R(\bm{p}_1,\bm{p}_2,\bm{p}_3)$ 
is a function of the momenta of the three incoming atoms.  
Galilean invariance implies that $R$ does not depend on the total 
momentum $\bm{p}_{\rm tot} = \bm{p}_1\, + \bm{p}_2\, + \bm{p}_3$.
It can therefore be expressed as a function 
$R(\bm{p}_{12},\bm{p}_{3,12})$
of a pair of Jacobi momenta $\bm{p}_{12}= \bm{p}_1 - \bm{p}_2$ 
and $\bm{p}_{3,12}= \bm{p}_3 - \frac12 (\bm{p}_1 + \bm{p}_2)$.
It is convenient to parameterize the Jacobi momenta in terms of 
6 orthogonal variables:
the collision energy $E = (3 p_{12}^2 + 4 p_{3,12}^2)/(12m)$, 
four Jacobi angles giving the orientations 
of the unit vectors $\hat{\bm{p}}_{12}$ and $\hat{\bm{p}}_{3,12}$,
and a hyperangle $\alpha_3= \arctan(\sqrt{3} p_{12}/(2 p_{3,12}))$. 
The integration element in these variables is
%----------------------
\begin{equation}
d^3 p_1 \, d^3 p_2 \, d^3 p_3
= \sqrt{3} \, m^3 E^2 dE \,  \sin^2(2 \alpha_3) d \alpha_3 \, 
d \Omega_{12} \, d \Omega_{3,12} \, d^3 p_{\rm tot}.
\label{dpp}
\end{equation}
%----------------------
%[e,hw,d]
We will refer to the average of a quantity over the Jacobi angles 
and over the hyperangle as the {\it hyperangular average}.
We denote the hyperangular average of the 3-body 
recombination rate $R(\bm{p}_{12},\bm{p}_{3,12})$ by $K(E)$.

In a gas of atoms with number density $n_A$,
the rate of decrease in the number density due to 3-body recombination
defines an experimentally measurable loss rate constant $L_3$: 
%----------------------
\begin{equation}
\frac{d \ }{d t} n_A 
= - L_3 \, n_A^3 \,.
\label{dn-lost}
\end{equation}
%----------------------
%[e,hw,l,d]
The event rate constant $\alpha$ for 3-body recombination
is defined so that the number of recombination 
events per unit volume and per unit time is $\alpha \, n_A^3$.  
If the number of atoms lost from the system per recombination event 
is $n_{\rm lost}$, the rate constant is $L_3 = n_{\rm lost} \, \alpha$.
The binding energy of the dimer is released through the  
kinetic energies of the recoiling atom and dimer. 
If their kinetic energies are large enough that the
atom and dimer both escape 
from the system and if the number density is low enough that 
subsequent collisions do not prevent them from escaping from the system, 
then $n_{\rm lost} = 3$.
On the other hand, if they both remain in the system 
and if we regard the dimer as a distinct chemical species 
with its own number density $n_D$, then $n_{\rm lost} = 2$.

In an ensemble of identical bosons with number density $n_A$
in thermal equilibrium at temperature $T$, the event rate 
per unit volume and per unit time is $\alpha n_A^3$, where 
%----------------------
\begin{eqnarray}
\alpha(T)= 
\frac{\int R(\bm{p}_{12},\bm{p}_{3,12}) \, 
n(E_1) n(E_2) n(E_3)  
	\, d^3p_1 d^3p_2 d^3p_3}
    {3! \, \int  n(E_1) n(E_2) n(E_3) \, d^3p_1 d^3p_2 d^3p_3} 
\label{Rfint}  
\end{eqnarray}
%----------------------
%[e,hw,l,d]
and $n(E)$ is the Bose-Einstein distribution:
$n(E) = [e^{(E - \mu)/(k_B T)} - 1]^{-1}$.
The chemical potential $\mu$ is determined by the condition 
$\int n(E_1) d^3p_1/(2 \pi)^3 = n_A$.
The factor of $1/3!$ in Eq.~(\ref{Rfint}) accounts for 
the indistinguishable nature of the three identical particles.
The denominator in Eq.~(\ref{Rfint}) can be written
$6 \, [\int n(E_1) \, d^3p_1]^3 = 6 (2 \pi)^9 n_A^3$.

The critical temperature $T_c$ for Bose-Einstein condensation is
given by $k_B T_c \approx 3.3 \, \hbar^2 n_A^{2/3}/m$.
If $T$ is significantly larger than $T_c$,
the product of the three  Bose-Einstein distributions 
in Eq.~(\ref{Rfint}) can be approximated by 
a Boltzmann distribution:
%----------------------
\begin{equation}
n(E_1) n(E_2) n(E_3) \approx e^{3 \mu/(k_B T)} \exp(-E_{\rm tot}/(k_B T)),
\end{equation}
%----------------------
%[e,hw,l,d]
where $E_{\rm tot} = E_1 + E_2 + E_3$ 
is the total energy of the three atoms.  
The total energy can be expressed as the sum of the center-of-mass energy 
and the collision energy $E$: 
$E_{\rm tot} = p_{\rm tot}^2/(6m) + E$.
The Gaussian integral over $\bm{p}_{\rm tot}$
cancels between the numerator and the denominator of Eq.~(\ref{Rfint}).
The effect of the integrals over the hyperangle and
the Jacobi angles is to replace $R(\bm{p}_{12},\bm{p}_{3,12})$ 
in the numerator by its hyperangular average $K(E)$.
Thus if $T$ is significantly larger than $T_c$, 
the event rate constant $\alpha(T)$ 
in Eq.~(\ref{Rfint}) reduces to 
a Boltzmann average over the collision energy $E$:
%----------------------
\begin{equation}
\alpha(T) \approx 
\frac{\int_0^\infty dE \, E^2 \, e^{-E/(k_B T)} \, K(E)}
    {6 \int_0^\infty dE \, E^2 \, e^{-E/(k_B T)}} \,.
\label{alpha-T}  
\end{equation}
%----------------------
%[e,hw,d]
The integral in the denominator 
can be evaluated analytically to give $2 (k_B T)^3$.
We will refer to temperatures and densities for which this 
approximation is valid as the {\it Boltzmann region}.
Note that the Boltzmann region can extend to arbitrarily low 
temperatures if the number density $n_A$ is arbitrarily small.
At $T=0$, the approximation in Eq.~(\ref{alpha-T})
reduces to $\alpha(0) = K(E=0)/6$.
However this applies only if the temperature is small 
enough that $K(E)$ can be approximated by $K(0)$
but still in the Boltzmann region.
At $T=0$, the atoms are in a Bose-Einstein condensate.
The correct result for the rate constant at $T=0$
is therefore $\alpha(0) \approx K(E=0)/36$.  
The extra factor of $1/3!$ comes from the 3 identical bosons 
being in the same quantum state.

If the scattering length $a$ is positive and large compared to the 
range of the interaction, one of the dimers 
that can be produced by the recombination process is the shallow dimer. 
Its binding energy in the zero-range limit is
%----------------------
\begin{eqnarray}
E_D = \hbar^2/(m a^2) \,.
\label{E-dimer}  
\end{eqnarray}
%----------------------
%[e,hw,d,l]
If there are deep dimers, they can be produced by the recombination
process for either sign of $a$.
The recombination rate can be decomposed 
into the contribution from the shallow dimer 
and the sum of the contributions from all the deep dimers:
%----------------------
\begin{equation}
K(E) =  K_{\rm shallow}(E) + K_{\rm deep}(E)  \,.
\label{K3}
\end{equation}
%----------------------
%[e,hw,l,d]
The 3-body recombination rate into the shallow dimer 
can be further decomposed into contributions from the channels 
in which the total orbital angular momentum of the three atoms
has definite quantum number $J$:
%----------------------
\begin{equation}
K_{\rm shallow}(E) = \sum_{J=0}^\infty K^{(J)}(E)  \,.
\label{K-shallow}
\end{equation}
%----------------------
%[hw,e,l,d]
The threshold behavior for each of the
angular momentum contributions to $K_{\rm shallow}(E)$
follows from a generalization of Wigner's threshold law \cite{EGS01}:
$K^{(J)}(E) \sim E^{\lambda_J}$,
where $\lambda_0 = 0$, $\lambda_1 = 3$, and $\lambda_J = J$
for $J \ge 2$.
At the scattering threshold $E=0$, only the $J=0$ term is nonzero.

The hyperangular average $K_{\rm shallow}(E)$ 
of the 3-body recombination rate is related in a simple way to the 
dimer-breakup cross section for the scattering of an atom
and the shallow dimer:
%----------------------
\begin{equation}
K_{\rm shallow}(E) = 
\frac{192 \sqrt{3} \pi \hbar^3 (E_D + E)}{m^2 E^2}
\sigma_{\rm breakup}(E).
\label{K-sigma}
\end{equation}
%----------------------
%[e,hw,d]
Thus in the Boltzmann region, the contribution  to the event rate 
constant in Eq.~(\ref{alpha-T}) from 3-body recombination into 
the shallow dimer is determined by the dimer-breakup cross section 
$\sigma_{\rm breakup}(E)$. 
Using the Optical Theorem, $\sigma_{\rm breakup}(E)$ can be determined 
from the phase shifts for elastic atom-dimer scattering.
We will show in Section~\ref{sec:deep} that in the case of a large 
scattering length, $K_{\rm deep}(E)$ is determined by the same universal
scaling functions that determine $K_{\rm shallow}(E)$.  Thus in the Boltzmann 
region, the rate constant for 3-body recombination 
is completely determined by the
elastic atom-dimer scattering amplitude.

\section{Universal scaling behavior}
\label{sec:nodeep}

In this section, we consider atoms with a large positive 
scattering length and no deep dimers,
so the only diatomic molecule is the shallow dimer
whose binding energy is given by Eq.~(\ref{E-dimer}).
The total 3-body recombination rate $K(E)$ 
is therefore equal to $K_{\rm shallow}(E)$, which is related 
to the dimer-breakup cross section by Eq.~(\ref{K-sigma}).
We summarize the constraints of universality
on the 3-body recombination rate for this case.

\subsection{Angular momentum and hyperangular decompositions}

The zero-range results for scattering rates, such as
$K_{\rm shallow}(E)$, are functions of the  
collision energy $E$, the scattering length $a$, and the 
Efimov parameter $\kappa_*$ defined by Eq.~(\ref{kappa-star}).  
An alternative Efimov parameter 
that is particularly convenient when considering 3-body 
recombination is $a_{*0}$, which is one of the values of the 
scattering length
at which $K_{\rm shallow}(E)$ vanishes at $E=0$.
This Efimov parameter differs from $\kappa_*^{-1}$ by a 
multiplicative factor that is known only to two digits of 
accuracy \cite{Braaten:2004rn}:
%----------------------
\begin{equation}
a_{*0}  \approx 0.32 \, \kappa_*^{-1} .
\label{a*0-kappa*}
\end{equation}
%----------------------
%[e,hw,d,l]
The dependence of scattering rates on the Efimov parameter 
is strongly constrained by the discrete scale invariance.
Their dependence on the energy $E$ can be conveniently 
expressed in terms of universal functions 
of a dimensionless scaling variable $x$ defined by
%----------------------
\begin{equation}
x = \left( m a^2 E/\hbar^2 \right)^{1/2} .
\label{x-def}
\end{equation}
%----------------------
%[hw,e,l,d]

Atom-dimer scattering states in the center-of-mass frame
can be labelled by the relative momentum $\bm{k}$ between 
the atom and the dimer or, equivalently, by the energy $E$
relative to the 3-atom threshold 
and the angular momentum quantum numbers $J$ and $M$. 
The $S$ matrix elements for
atom-dimer elastic scattering are functions of $E$ that are 
independent of $M$ and diagonal in $J$.  They can be expressed 
in terms of the phase shifts $\delta_{AD}^{(J)}(E)$ 
for atom-dimer elastic scattering:
%----------------------
\begin{equation}
S_{AD,AD}^{(J)}(E) = e^{2 i \delta_{AD}^{(J)}(E) } .
\label{S-deltaAD}
\end{equation}
%----------------------
%[e,hw,l,d]
The phase shifts $\delta_{AD}^{(J)}(E)$ are real below the 3-atom 
threshold and have a positive imaginary part for $E>0$.

Three-atom scattering states in the center-of-mass frame
can be labelled by the Jacobi momenta $\bm{p}_{12}$ and 
$\bm{p}_{3,12}$.  One can define subsystem angular momenta 
associated with rotations of the unit vectors $\hat{\bm{p}}_{12}$ and 
$\hat{\bm{p}}_{3,12}$.  We denote the associated angular 
momentum quantum numbers by $\ell_x$, $m_x$ and by $\ell_y$, $m_y$,
respectively.
An equivalent set of quantum numbers are $\ell_x$, $\ell_y$, 
and the total orbital angular momentum quantum numbers $J$ and $M$. 
The possible values of $J$ are integers 
ranging from $|\ell_x -\ell_y|$ to $\ell_x + \ell_y$.
The range of the hyperangle $\alpha_3$ is from 0 to $\frac12 \pi$.
Since this range is compact, we can expand functions of 
$\alpha_3$ in terms of orthogonal functions relative to the 
weight factor $\sin^2 (2 \alpha_3)$ that are
labelled by an index $n_3 = 0,1,2, \ldots$.
A convenient set of variables for the 3-atom scattering  states 
is the collision energy $E$ and five discrete variables:
$J$, $M$, $\ell_x$, $\ell_y$, and the hyperangular index $n_3$.  
Since $\ell_x$, $\ell_y$, and $n_3$ 
are a denumerable set, we will denote them collectively 
by a single integer $n$ that takes values $3, 4, 5, \ldots$.  
The reason for choosing this peculiar set of 
values will become clear in Section~\ref{sec:scalingJ=0}.
The S-matrix elements for 3-body recombination can be expressed 
as functions $S^{(J,n)}_{AAA,AD}(E)$ 
of the collision energy with index $n$.
The contribution from angular momentum $J$  
to the hyperangular average of the 
3-body recombination rate can be expressed as
%----------------------
\begin{equation}
K^{(J)}(E) = \frac{144 \sqrt{3} \pi^2 \hbar^5 (2J+1)}{m^3 E^2}
\sum_{n=3}^\infty \left| S_{AAA,AD}^{(J,n)}(E) \right|^2.
\label{Kshallow-S}
\end{equation}
%----------------------
%[hw,e,d,l]
The hyperangular average is implemented by the sum over $n$.
The unitarity of the S-matrix in the angular momentum $J$ sector 
implies
%----------------------
\begin{equation}
|S_{AD,AD}^{(J)}(E)|^2 
+ \sum_{n=3}^\infty |S_{AD,AAA}^{(J,n)}(E)|^2 = 1 .
\label{unitarity:0}
\end{equation}
%----------------------
%[hw,e,l,d]
This can be used together with Eq.~(\ref{S-deltaAD}) to express
the 3-body recombination rate in Eq.~(\ref{Kshallow-S})
in terms of the phase shifts $\delta_{AD}^{(J)}(E)$
for elastic atom-dimer scattering:
%----------------------
\begin{equation}
K^{(J)}(E) = \frac{144 \sqrt{3} \pi^2 \hbar^5 (2J+1)}{m^3 E^2}
\left( 1 - \Big| e^{2 i \delta_{AD}^{(J)}(E)} \Big|^2 \right) .
\label{K-deltaAD}
\end{equation}
%----------------------
%[hw,e,l,d]

\subsection{Scaling behavior for $\bm{J \geq 1}$}

In the sector with angular momentum quantum number $J \geq 1$,
the scattering rates do not depend on $\kappa_*$.
The constraints of universality are 
therefore particularly simple for $J \ge 1$. 
The atom-dimer phase shifts $\delta_{AD}^{(J)}(E)$
defined by Eq.~(\ref{S-deltaAD}) are universal functions 
of the scaling variable $x$ defined in Eq.~(\ref{x-def}).
Below the 3-atom threshold, these scaling functions are real.
Above the 3-atom threshold, they are complex.

The contribution to the 3-body recombination rate into the shallow dimer
from angular momentum $J$ has the form 
%----------------------
\begin{equation}
K^{(J)}(E) = \frac{144 \sqrt{3} \pi^2 (2J+1) f_J(x)}{x^4} \, 
\frac{\hbar a^4}{m} \,,
\label{alpha-J:uni}
\end{equation}
%----------------------
%[hw,e,l,d]
where $f_J(x)$ is a real-valued scaling function:
%----------------------
\begin{equation}
f_J(x) = 1 - \exp \big(-4 \, {\rm Im} \, \delta_{AD}^{(J)}(E) \big)  \,.
\label{fJ-delta}
\end{equation}
%----------------------
%[hw,e,l,d]
As $x \to 0$, the leading powers of $x$ are determined by 
Wigner's threshold law \cite{EGS01}:
$f_J(x) \sim x^{2 \lambda_J+4}$,
where $\lambda_1 = 3$ and $\lambda_J = J$  for $J \ge 2$.

\subsection{Discrete scaling behavior for $\bm{J = 0}$}
\label{sec:scalingJ=0}

The zero-range results for $J=0$ are more intricate than for $J \geq 1$,
because they depend not only on $E$ and $a$,
but also on $\kappa_*$ or, equivalently, $a_{*0}$.
General scaling formulas for scattering rates in the 3-atom sector 
can be derived from Efimov's Radial Law.  The Radial Law 
expresses the S-matrix elements for low-energy scattering processes 
in the $J=0$ channel of the 3-atom sector
in terms of universal functions of  
the scaling variable $x$ defined in Eq.~(\ref{x-def}).
The S-matrix elements for atom-dimer elastic scattering and 
for 3-body recombination from 3-atom states labelled $(J=0,n)$
have the form \cite{Braaten:2004rn}:
%----------------------
\begin{subequations}
\begin{eqnarray}
S_{AD,AD}^{(J=0)}(E) & = & s_{22}(x)
+ \frac{s_{21}(x)^2 e^{2i \theta_{*0}}}
      {1 - s_{11}(x) e^{2i \theta_{*0}}}  \,,
\label{RL:AD}
\\
S_{AD,AAA}^{(J=0,n)}(E) & = & s_{2n}(x) 
+ \frac{s_{21}(x) s_{1n}(x) e^{2i \theta_{*0}}}
      {1 - s_{11}(x) e^{2i \theta_{*0}}}  \,,
\label{RL:ADAAA}
\end{eqnarray}
\label{RL}
\end{subequations}
%----------------------
%[e,hw,l,d]
where the angle $\theta_{*0}$ is
%----------------------
\begin{equation}
\theta_{*0} = s_0 \ln (a/a_{*0}) .
\label{theta*}
\end{equation}
%----------------------
%[e,hw,l,d]
The functions $s_{ij}(x)$ are entries of 
an infinite-dimensional  
symmetric unitary matrix that depends only on the scaling 
variable $x$ defined in Eq.~(\ref{x-def}). 
The unitarity of the infinite-dimensional matrix $s$ 
can be expressed as
%----------------------
\begin{equation}
s_{1i}^* s_{1j} + s_{2i}^* s_{2j} 
+ \sum_{n=3}^\infty s_{ni}^* s_{nj} = \delta_{ij} .
\label{s:unitary}
\end{equation}
%----------------------
%[e,hw,d,l]
This condition implies the unitarity of the 
physical S-matrix for the $J=0$ sector.  
For example, one can use Eq.~(\ref{s:unitary}) to verify
that the unitarity condition in Eq.~(\ref{unitarity:0})
for $J=0$ is automatically satisfied if the S-matrix elements 
are given by the expressions in Eqs.~(\ref{RL}).

Efimov's Radial Law has a simple interpretation
in terms of the adiabatic hyperspherical representation 
of the 3-atom Schr\"odinger equation.
The hyperradius $R = [(r_{12}^2 + r_{23}^2 + r_{31}^2)/3]^{1/2}$
is the root-mean-square separation of the three atoms. 
We take the scattering length $a$ to be much larger than the 
absolute value of the effective range $r_s$: $a \gg |r_s|$.
In this case, there are four important regions of $R$ for 3-atom 
configurations with energies near the scattering threshold:
\begin{itemize}

\item the {\it asymptotic} region $R \gg a$,

\item the {\it scaling} region $R \sim a$,

\item the {\it scale-invariant} region $|r_s| \ll R \ll a$,

\item the {\it short-distance} region $R \sim |r_s|$.

\end{itemize}
Efimov's radial law reflects the fact that there is only one 
adiabatic hyperspherical potential that is attractive 
in the scale-invariant region of $R$.
It is only through this adiabatic hyperspherical potential 
that 3-atom configurations with energies near the scattering 
threshold can reach the short-distance region of $R$.
The outgoing and incoming asymptotic states with respect 
to the scaling region are
\begin{enumerate}

\item incoming and outgoing hyperradial waves
in the scale-invariant region $|r_s| \ll R \ll a$ of the lowest 
adiabatic hyperspherical potential.
We label these asymptotic states by the index 1.

\item outgoing or incoming atom-dimer scattering states
with angular momentum quantum number $J=0$.
We label these asymptotic states $AD$ or simply by the index 2.

\item outgoing or incoming 3-atom scattering states
with $J=0$.
We label these asymptotic states $AAA$ by the index 
$n=3, 4, 5, \ldots$ that specifies $\ell_x$, $\ell_y$, and $n_3$.

\end{enumerate}
The evolution of the 3-atom wavefunction through 
the scaling region is described by a  
unitary matrix $s_{ij}$ whose indices correspond to the 
asymptotic states relative to the scaling region 
that were enumerated above.  Time-reversal invariance implies 
that the matrix $s$ is symmetric.  

The expressions for the S-matrix elements in Eq.~(\ref{RL})
have simple interpretations.  The factor $e^{2i \theta_{*0}}$
in Eqs.~(\ref{RL}) is the phase shift due to reflection
of a hyperradial wave from the short-distance region.
The S-matrix elements in Eqs.~(\ref{RL}) can be
expanded as a power series in $e^{2i \theta_{*0}}$
with each term corresponding to a different pathway between 
the incoming and outgoing scattering states.
The term with the factor $(e^{2i \theta_{*0}})^n$ is the amplitude
for a pathway that includes $n$ reflections from the 
short-distance region.  In the S-matrix element 
$S_{AD,AD}^{(J=0)}(E)$ in Eq.~(\ref{RL:AD}), 
the leading term $s_{22}$ is the amplitude for the incoming atom-dimer 
scattering state to be reflected from the scaling region.
The second term $s_{21} e^{2i \theta_{*0}} s_{12}$ is the 
amplitude for the atom-dimer scattering state to be transmitted 
through the scaling region to an incoming hyperradial wave, 
which is reflected 
from the short-distance region into an outgoing hyperradial wave, 
which is then transmitted through the scaling region to an outgoing 
atom-dimer scattering state.  The factors of 
$e^{2i \theta_{*0}} s_{11}$ from expanding the denominator are
the amplitudes for the hyperradial wave to be reflected from
the short-distance region and then reflected from the scaling region.

The contribution to the hyperangular average of the 
3-body recombination rate from angular momentum $J=0$ can be obtained 
by using the unitarity condition in Eq.~(\ref{unitarity:0})
to eliminate the sum over $n$ in Eq.~(\ref{Kshallow-S})
and then inserting the expression for the S-matrix element 
in Eq.~(\ref{RL:AD}):
%----------------------
\begin{equation}
K^{(0)}(E) = 
\frac{144\sqrt{3} \pi^2}{x^4}
\left(1 - \left| s_{22}(x) 
+ \frac{s_{12}(x)^2 e^{2 i\theta_{*0}}}
      {1 - s_{11}(x) e^{2 i \theta_{*0}}}
\right|^2 \right)  \frac{\hbar a^4}{m} \,.
\label{sigbr:rl}
\end{equation}
%----------------------
%[e,hw,l,d]

\subsection{Analytic results near threshold}

Macek, Ovchinnikov, and Gasaneo have derived some remarkable analytic 
results for the 3-body problem of identical bosons with large
scattering length and energies near the 3-atom threshold \cite{MOG05,MOG06}.
In Ref.~\cite{MOG05}, they derived an analytic result 
for the S-wave phase shift $\delta_{AD}^{(J=0)}(0)$ for atom-dimer 
elastic scattering at the breakup threshold $E=0$.
They obtained  analytic results for the values 
$s_{11}(0)$ and $s_{12}(0)$
of the universal scaling functions at this threshold.
In Ref.~\cite{MOG06}, they extended their analytic solution for 
$\delta_{AD}^{(J=0)}(E)$ to the first nontrivial order in $E$.
They used this result to derive an analytic expression for the 
3-body recombination rate $K_{\rm shallow}(0)$ at 
the 3-atom threshold.  

An analytic expression for the atom-dimer 
S-wave phase shift $\delta^{(0)}_{AD}(E)$
near the dimer-breakup threshold is given in Ref.~\cite{MOG06}.
Using the first equality in Eq.~(54) of Ref.~\cite{MOG06}
(to avoid typographical errors in the last line of Eq.~(54)), 
the S-matrix element at energy $E=x^2 E_D$ 
can be expressed in the form
%----------------------
\beq
\exp \big( 2i\delta_{AD}^{(0)}(E) \big) =
\frac{e^{2i [\delta_\infty(x) - \Delta(x)]}}{1 + e^{-2\pi s_0} e^{-2i\Delta(x)}}
\left[ 1 + e^{-2\pi s_0} e^{2i\Delta(x)}
+ e^{- \pi s_0} A(x) \left( e^{2i\Delta(x)} - 1 \right) \right]~,
\label{deltaAD-analytic}
\eeq
%----------------------
%[e,l,hw,d]
where 
%----------------------
\beq
\Delta(x) = \delta_0(x) - s_0 \ln(a/R_0).
\label{theta*0-delta0}
\eeq
%----------------------
%[e,hw,l,d]
The variable $R_0$ is a matching point at short distances
where the 3-body wavefunction vanishes.
(The function $\Delta(x)$ was denoted by $\Delta(R_0)$ in Ref.~\cite{MOG06}.)
The function $A(x)$ and the energy dependent phases $\delta_\infty(x)$
and $\delta_0(x)$ are universal functions of the scaling variable $x$.
By replacing the 
expression inside the absolute value signs in Eq.~(\ref{sigbr:rl})
by the phase shift in Eq.~(\ref{deltaAD-analytic}), 
the 3-body recombination rate can be expressed in the form
%-----------------
\begin{eqnarray}
K^{(0)}(E) = 
288 \sqrt{3} \pi^2 \frac{A(x)}{x^4}
\left( 1 - \frac{A(x)}{2\sinh(\pi s_0)} \right) 
\frac{\sinh(\pi s_0) \sin^2 \Delta(x)}
    {\sinh^2(\pi s_0) + \cos^2 \Delta(x)} \,
\frac{\hbar a^4}{m} \,.
\label{K0-MOG}
\end{eqnarray}
%----------------------
%[e,hw]

To determine the 3-body recombination rate at threshold,
we need the limiting behavior of the functions
$A(x)$ and $\delta_0(x)$ as $x \to 0$.
The values of these phases at $x=0$
were calculated numerically in Ref.~\cite{MOG05}:
%-----------------
\begin{subequations}
\begin{eqnarray}
\delta_\infty(0) &=& 1.736 \,,
\\
\delta_0(0) &=& 1.588 \,.
\end{eqnarray}
\label{delta-inf0}
\end{subequations}
%----------------------
%
%[e,d,hw,l]
The leading term in $A(x)$ as $x \to 0$, which is proportional to $x^4$,
was calculated analytically in Ref.~\cite{MOG06}:
%----------------------
\beq
A(x) \longrightarrow 
\frac{8(4\pi-3\sqrt{3})}{3\sqrt{3}\sinh(\pi s_0)} x^4 ,
\label{A-x}
\eeq
%----------------------
%[e,hw,l,d]
The recombination rate in Eq.~(\ref{K0-MOG}) vanishes at the threshold 
if $\Delta(0)=0$.  Thus the expression for $\Delta(x)$
in Eq.~(\ref{theta*0-delta0}) can be written
%----------------------
\beq
\Delta(x) = \delta_0(x) - \delta_0(0) - s_0 \ln(a/a_{*0}) ,
\label{Delta-theta*0}
\eeq
%----------------------
%[e,l,hw,d]
where $a_{*0}$ is a value of the scattering length for which the 3-body
recombination rate vanishes at threshold.
The expression for the S-wave atom-dimer phase shift in 
Eq.~(\ref{deltaAD-analytic}) reduces at the 3-atom 
threshold to
%----------------------
\begin{eqnarray}
\delta_{AD}^{(0)}(E=0) = 
\delta_\infty(0) 
+ \arctan \big( \tanh (\pi s_0) \tan \theta_{*0} \big) \,.
\label{deltaAD0-analytic}
\end{eqnarray}
%----------------------
%[e,l,hw,d]
The expression given in Ref.~\cite{MOG05} is equivalent but less compact.
The analytic result of Macek, Ovchinnikov, and Gasaneo 
for the 3-body recombination rate at the 3-atom threshold 
\cite{MOG06} can be obtained by taking the limit $x \to 0$ in 
Eq.~(\ref{K0-MOG}):
%-----------------
\begin{eqnarray}
K^{(0)}(E=0) = 
\frac{768 \pi^2 (4 \pi - 3 \sqrt{3}) \sin^2[s_0 \ln(a/a_{*0})]}
    {\sinh^2(\pi s_0) + \cos^2[s_0 \ln(a/a_{*0})]} \,
\frac{\hbar a^4}{m} \,.
\label{K0-analytic}
\end{eqnarray}
%----------------------
%[hw,e,l,d]
This result was also derived independently by Petrov \cite{Petrov-octs}.
For fixed $a$, the rate in Eq.~(\ref{K0-analytic})
is a log-periodic function of $a_{*0}$ that oscillates between zero and 
a maximum value
%-----------------
\begin{eqnarray}
K_{\rm max} = 
6 C_{\rm max} \hbar a^4/m \,,
\label{Kmax}
\end{eqnarray}
%----------------------
%[hw,d,e,l]
where $C_{\rm max}$ is
%-----------------
\begin{eqnarray}
C_{\rm max} = 
\frac{128 \pi^2 (4 \pi - 3 \sqrt{3})}{\sinh^2(\pi s_0)} \,.
\label{K-max}
\end{eqnarray}
%----------------------
%[hw,e,d,l]
Its numerical value is $C_{\rm max} \approx 67.1$.
The expression in Eq.~(\ref{K0-analytic}) has zeroes when 
$a$ is $( e^{\pi/s_0})^n \, a_{*0}$, where $n$ is an integer.
The maxima of 
$K^{(0)}(0)$ in Eq.~(\ref{alpha-sh}) occur when 
$a$ is $( e^{\pi/s_0} )^n \, 14.3 \, a_{*0}$.
Since $\sinh^2(\pi s_0) \approx 139$ is so large, the expression for
$K^{(0)}(0)$ in Eq.~(\ref{K0-analytic}) can be approximated 
with an error of less than 1\% of $6 C_{\rm max}\hbar a^4/m$ by
%----------------------
\begin{equation}
K^{(0)}(E=0) \approx 
6 C_{\rm max} \sin^2 [s_0 \ln (a/a_{*0}) ] \, \hbar a^4/m \,.
\label{alpha-sh}
\end{equation}
%----------------------
%[hw,e,l,d]
This approximate functional form of the rate constant was deduced
independently in Refs.~\cite{NM-99,EGB-99,Bedaque:2000ft}.

By comparing the expression for the S-matrix element
in Eq.~(\ref{deltaAD-analytic}) with the Efimov's Radial Law
in Eq.~(\ref{RL:AD}), 
we can determine the universal scaling functions
$s_{11}(x)$, $s_{12}(x)$, and $s_{22}(x)$ 
in the approximation of Ref.~\cite{MOG06}:
%----------------------
\begin{subequations}
\bea
s_{11}(x) &\approx& -  e^{-2\pi s_0} e^{-2i[\delta_0(x) - \delta_0(0)]}~,
\label{s11:x->0}
\\
s_{12}(x) &\approx& 
e^{i [\delta_\infty(x) - \delta_0(x) + \delta_0(0)]} \sqrt{1-e^{-4\pi s_0}} 
\left[ 1 - \frac{A(x)}{2 \sinh(\pi s_0)} \right]^{1/2} ,
\label{s12:x->0}
\\
s_{22}(x) &\approx& 
e^{2 i \delta_\infty(x)} e^{-2\pi s_0} 
\left[ 1 + e^{\pi s_0} A(x) \right] .
\label{s22:x->0}
\eea
\label{sij:x->0}
\end{subequations}
%----------------------
%[e,l,hw]
The limiting behavior of the functions
$\delta_\infty(x)$, $\delta_0(x)$, and $A(x)$ as $x \to 0$, 
is given in Eqs.~(\ref{delta-inf0}) and (\ref{A-x}).
At $x=0$, we obtain $s_{11}(0)= - e^{-2 \pi s_0}$ and 
$s_{12}(0)= e^{i \delta_\infty(0)} (1-e^{-4 \pi s_0})^{1/2}$.
These differ from the values deduced in Ref.~\cite{MOG05}
because a phase
$e^{2i \delta_0(0)}$ has been absorbed into $e^{2 i \theta_{*0}}$.
They also differ by a minus sign in $s_{11}(0)$.
For small $x$, the leading terms in $s_{12}(x) - s_{12}(0)$
and $s_{22}(x) - s_{22}(0)$ scale like $x^4$.
The leading terms in $s_{11}(x) - s_{11}(0)$ 
scale like a higher power of $x$.

\section{Universal Scaling Functions}
\label{sec:unisf}

In this Section, we calculate the universal scaling functions 
associated with atom-dimer elastic scattering.
These functions are then used to calculate the hyperangular average 
of the 3-body recombination rate as a function of the collision energy.

\subsection{STM Equation}

Three-body observables for systems with a large scattering length can be
calculated by solving an integral equation called the
Skorniakov--Ter-Martirosian (STM) equation \cite{STM57}. 
In momentum space, the STM
equation can be expressed in the form \cite{Braaten:2004rn}
%----------------------
\begin{eqnarray}
 {\mathcal A} ({\bm p}, {\bm k}; E)
&=& 
- \frac{16 \pi/a} 
      {mE - (p^2 + {\bm p} \cdot {\bm k} + k^2) +i \epsilon}
\nonumber
 \\
&& 
- \int \frac{d^3q}{(2 \pi)^3}
\frac{8 \pi}{mE - (p^2 + {\bm p} \cdot {\bm q} + q^2) + i \epsilon} \,
\frac{{\mathcal A} ({\bm q}, {\bm k}; E)}
{- 1/a + \sqrt{- mE + 3q^2 /4 -i \epsilon} } \,.
\label{BhvK:general}
\end{eqnarray}
%----------------------
%[d,e,hw,l]
The solution is, up to a normalization constant, the amplitude for an atom
with momentum $\bm{p}$ and energy $p^2/(2m)$ and a pair of atoms with
total momentum $- \bm{p}$ and total energy $E-p^2/(2m)$ to evolve into an
atom with momentum $\bm{k}$ and energy $k^2/(2m)$ and a pair of atoms
with total momentum $-\bm{k}$ and total energy $E - k^2/(2m)$. The STM
amplitude can be resolved into the contributions from channels with
orbital angular momentum quantum number $J$:  
%
%----------------------
\begin{equation}
\mathcal{A}_{J} (p, k; E) =  
\mbox{$\frac12$} \int_{-1}^1 d y \, 
P_{J}(y) \mathcal{A} (\bm{p}, \bm{k}; E)\,,
\end{equation}
%----------------------
%[hw,l]
where $y=\bm{p}\cdot \bm{k}/pk$ and $P_{J}(y)$ is a Legendre 
polynomial.  

In a specific angular momentum channel, the STM equation 
in Eq.~(\ref{BhvK:general}) reduces to an integral equation
with a single integration variable $q$.
In the case $J = 0$, the behavior of the integral is sufficiently
singular at large $q$ that it is necessary to impose an ultraviolet
cutoff $q < \Lambda$, where $\Lambda$ is much greater than $p$, $k$ and
$(m|E|)^{1/2}$. The STM equation for $J=0$ reduces to
%----------------------
\begin{eqnarray}
{\mathcal A}_0 (p, k; E, \Lambda)  &=& 
\frac{8 \pi}{a p k} 
\ln \frac{p^2 + pk + k^2 -mE -i \epsilon}
       {p^2 - pk + k^2 - mE - i \epsilon}
\nonumber
\\
&+&
\frac{2}{\pi} \int_0^\Lambda dq \, \frac{q}{p} 
\ln \frac{p^2 +pq + q^2 - mE - i \epsilon}
             {p^2 - pq + q^2 -mE -i \epsilon} \,
\frac{ {\mathcal A}_0 (q, k; E, \Lambda)}
{- 1/a + \sqrt{3q^2/4 -mE -i \epsilon} }\,.
\label{BHvK}
\end{eqnarray}
%----------------------
%[d,e,hw,l]
It can be shown that
changing the ultraviolet cutoff $\Lambda$ corresponds to
changing the Efimov parameter $\kappa_*$
\cite{Bedaque:1998kg,Bedaque:1998km,Hammer:2000nf}. 
The values of  $\Lambda$ and $\kappa_*$ differ simply
by a multiplicative constant:
%----------------------
\beq
\ln(\Lambda/\kappa_*) \approx 1.74 \mod (\pi/s_0)\,.
\eeq
%----------------------
%[hw,l]
The solutions of Eq.~(\ref{BHvK}) are log-periodic functions of 
$\Lambda$.  If $\Lambda$ is increased by a factor of 
$e^{\pi/s_0} \approx 22.7$, 
it corresponds to the same value of $\kappa_*$.
The S-wave atom-dimer phase shift
$\delta^{(0)}_{AD}(E)$ can be obtained from the amplitude with both
momentum arguments $p$ and $k$ set equal to 
$k_E = (\frac 4 3 m (E_D+E))^{1/2}$:
%
%----------------------
\begin{equation}
\mathcal{A}_{0} (k_E, k_E; E, \Lambda) = 
\frac {3\pi} {k_E \cot \delta^{(0)}_{AD} (E) -ik_E}\,.
\label{A-deltaAD}
\end{equation}
%----------------------
%
%[hw,l,d,e]
This phase shift depends not only on the parameter
$a$ but also on the parameter $\kappa_*$ through the dependence 
of the amplitude on $\Lambda$.

In the case $J \geq 1$, no ultraviolet cutoff is required and 
the partial wave projection of the STM equation 
leads to (see, e.g. Ref.~\cite{Griesshammer:2005ga})
%----------------------
\begin{eqnarray}
{\mathcal A}_J (p, k; E)  &=& 
\frac{16\pi}{a p k}\, (-1)^J\, 
Q_J \bigg( \frac{p^2 + k^2 -mE -i \epsilon}{pk} \bigg)
\nonumber
\\
&&
+ \frac{4}{\pi} \int_0^\infty dq \, \frac{q}{p}\,
Q_J \Big( \frac{p^2 + q^2 - mE - i \epsilon}{pq} \Big) \,
\frac{ \, (-1)^J\, {\mathcal A}_J (q, k; E)}
{- 1/a + \sqrt{3q^2/4 -mE -i \epsilon} } \, ,
\label{BHvKJ}
\end{eqnarray}
%----------------------
%[hw,l]
where 
%
%----------------------
\beq
Q_J(z) =\frac{1}{2}\int_{-1}^1 dx \frac{P_J(x)}{z-x} 
\eeq
%----------------------
%
%[hw,l,e]
is a Legendre function of the second kind.
The atom-dimer phase shift $\delta^{(J)}_{AD}(E)$ can be 
obtained from 
%
%----------------------
\begin{equation}
\mathcal{A}_{J} (k_E, k_E; E) = 
\frac {3\pi} {k_E \cot \delta^{(J)}_{AD} (E) -ik_E}\,,
\qquad \mbox{ for }\quad J\geq 1\,.
\label{A-deltaAD-J}
\end{equation}
%----------------------
%
%[hw,l]
This phase shift is a function of the dimensionless 
variable $ma^2E/\hbar^2$.

\subsection{Universal scaling functions for $\bm{J \ge 1}$}

%%%%%%%%%%%%%%%%%%%%%%%%%%%%%%%%%%%%%%%%%%%%%%%%%%
\begin{figure}[htb]
\includegraphics*[width=12cm,angle=0,clip=true]{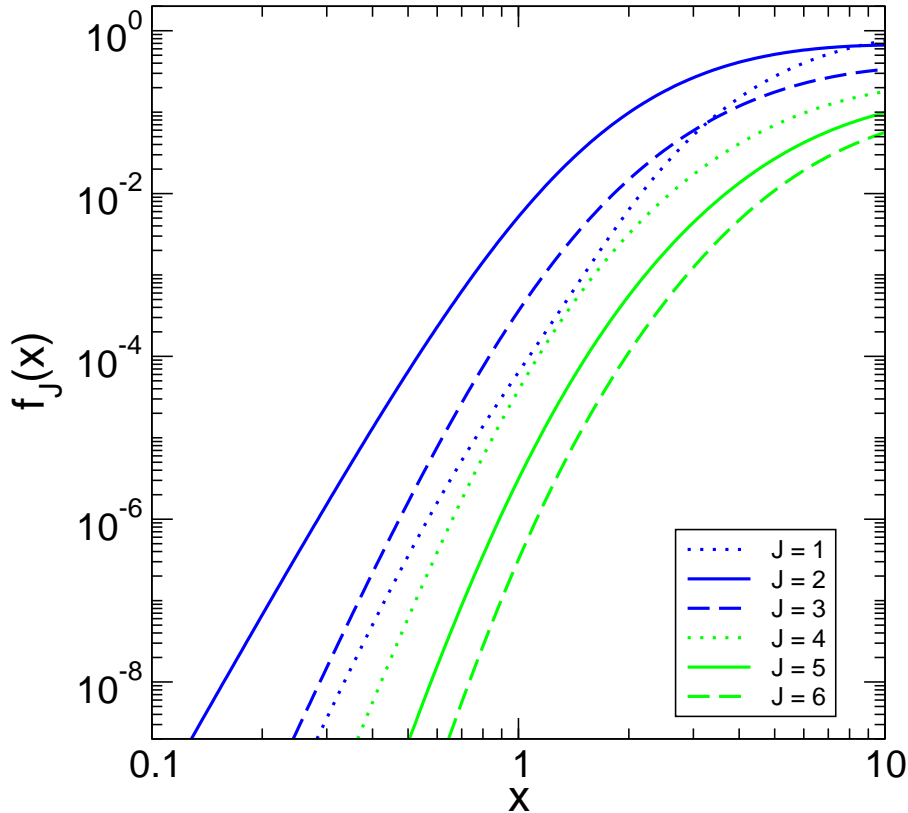} 
\vspace*{0.0cm}
\caption{The  universal scaling functions 
$f_J(x) = 1 - \exp (-4 {\rm Im} \delta_{AD}^{(J)}(E) )$
for $J=1, \ldots, 6$ as functions of 
the scaling variable $x$.}
\label{fig:fij-x}
\end{figure}
%%%%%%%%%%%%%%%%%%%%%%%%%%%%%%%%%%%%%%%%%%%%%%%%%%

We calculate the atom-dimer phase shifts $\delta^{(J)}_{AD}(E)$
for $J = 1,2,\ldots,6$ as functions of the energy $E$ from
$10^{-2} E_D$ to $10^2 E_D$. 
The scaling variable $x$ defined in Eq.~(\ref{x-def}) 
then ranges from 0.1 to 10.
For each energy $E$, we solve
the STM equation in Eq.~(\ref{BHvKJ})
and determine the phase shift $\delta^{(J)}_{AD}(E)$
using Eq.~(\ref{A-deltaAD}).  The universal scaling function $f_J(x)$ 
defined in Eq.~(\ref{fJ-delta}) is then determined from the 
imaginary part of the phase shift.
Our numerical method loses accuracy for $x$ smaller than values 
that range from 0.1 for $J=2$ to $x = 0.45$ for $J=6$.
To determine $f_J(x)$ for smaller values of $x$, we fit our results 
for the lowest 10 accurate points to the form
%----------------------
\beq
f_J(x) = a_J x ^{2 \lambda_J+4} + b_J x ^{2 \lambda_J + 6},
\label{fJ-x}
\eeq
%----------------------
%[e,d]
where $\lambda_1=3$ and $\lambda_J=J$ for $J \ge 2$.
We use the formula in Eq.~(\ref{fJ-x}) to extrapolate 
to small values of $x$ where our numerical method loses accuracy.
Our results are shown in Fig.~\ref{fig:fij-x}.
For $J \ge 2$, $f_J(x)$ is a decreasing function of $J$.
This pattern is broken by $f_1(x)$, which is smaller than 
$f_2(x)$ and $f_3(x)$ at small $x$ 
but eventually becomes larger than $f_2(x)$ around $x = 9$.

\subsection{Universal scaling functions for $\bm{J=0}$}

%%%%%%%%%%%%%%%%%%%%%%%%%%%%%%%%%%%%%%%%%%%%%%%%%%
\begin{figure}[htb]
\includegraphics*[width=6.6cm,angle=0,clip=true]{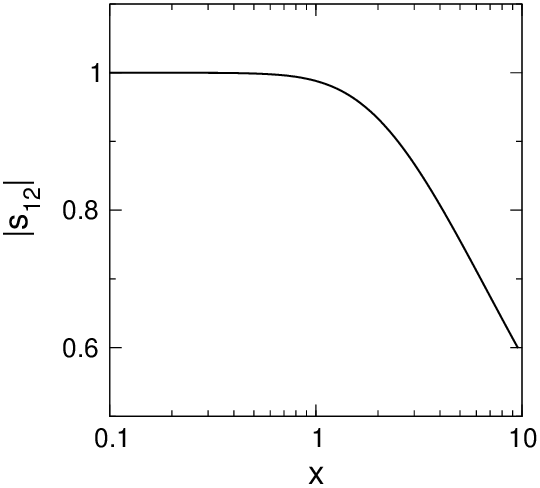} \hspace{1cm}
\includegraphics*[width=6.6cm,angle=0,clip=true]{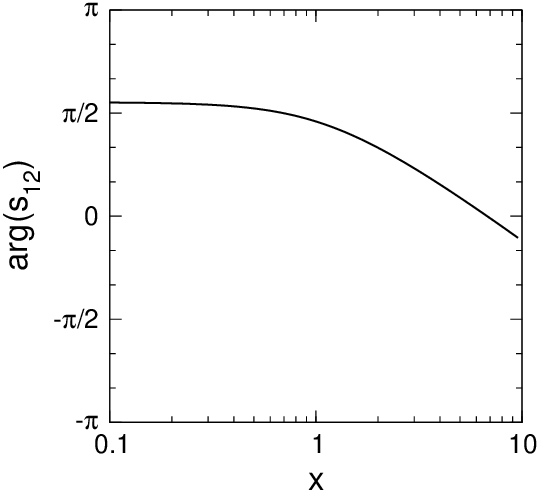}
\vspace{1cm}
\includegraphics*[width=6.6cm,angle=0,clip=true]{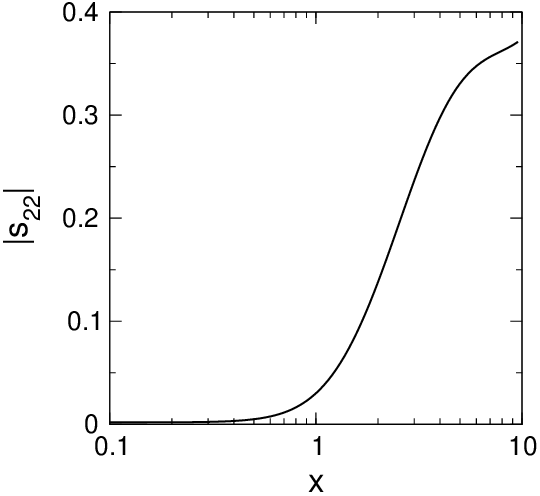} \hspace{1cm}
\includegraphics*[width=6.6cm,angle=0,clip=true]{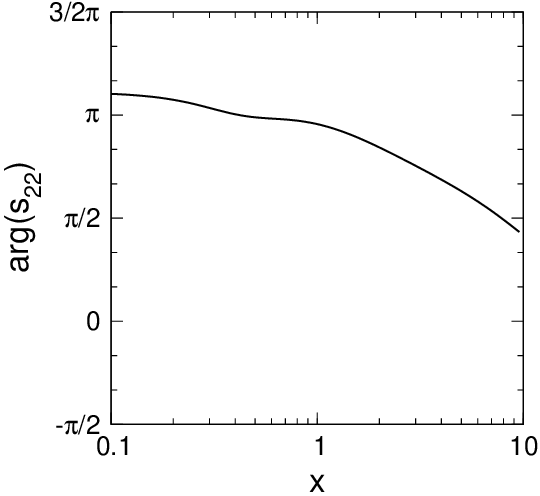}
\vspace{1cm}
\includegraphics*[width=6.6cm,angle=0,clip=true]{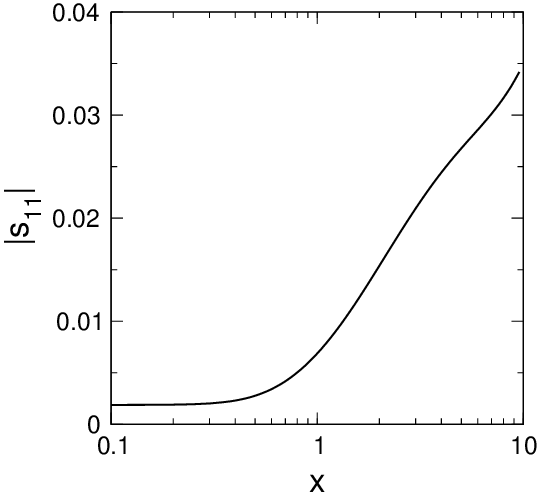} \hspace{1cm}
\includegraphics*[width=6.6cm,angle=0,clip=true]{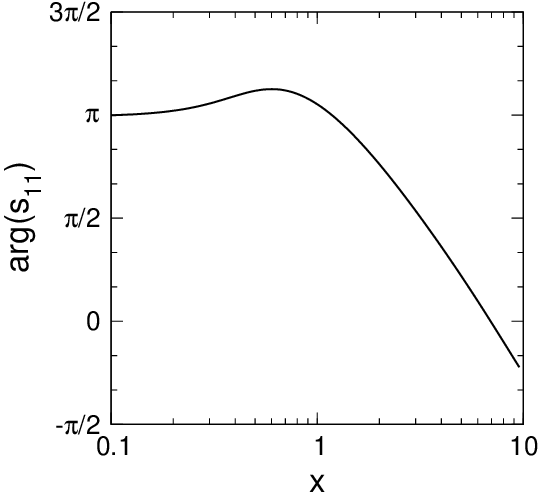}
\vspace*{0.0cm}
\caption{The  universal scaling functions 
$s_{12}(x)$ (upper panels), $s_{22}(x)$ (middle panels), 
and $s_{11}(x)$ (lower panels) as functions of $x$. 
The modulus $|s_{ij}(x)|$ and the phase $\arg s_{ij}(x)$
of the function are shown in the left panel 
and the right panel, respectively.}
\label{fig:sij-x}
\end{figure}
%%%%%%%%%%%%%%%%%%%%%%%%%%%%%%%%%%%%%%%%%%%%%%%%%%

We calculate the atom-dimer phase shift $\delta^{(0)}_{AD} (E)$ 
by solving the STM
equation for $J=0$ with an ultraviolet cutoff $\Lambda$, 
which is given in Eq.~(\ref{BHvK}), for energies $E$ 
ranging from $10^{-2} E_D$ to $10^2 E_D$.
We choose values of the
ultraviolet cutoff $\Lambda$ that are equally spaced on a log scale and
cover three quarters of a discrete scaling cycle 
from $\Lambda_0$ to $10.4\Lambda_0$.
The smallest cutoff $\Lambda_0$ is chosen to
be much greater than $1/a$ and $(m|E|/\hbar^2)^{1/2}$ 
and large enough that the calculated phaseshift is 
unchanged if $\Lambda$ is increased by a factor of 22.7.
For each energy $E$, we calculate the phase shift as a function of 
$\log (a \Lambda)$ and fit it to the universal formula 
%----------------------
\begin{eqnarray}
\exp \left( 2 i \delta_{AD}^{(J=0)}(E) \right)  &=& 
s_{22}(x)  
+ \frac{s_{12}(x)^2 \exp[2is_0 \ln(a/a_{*0})]}{1 - 
  s_{11}(x) \exp[2is_0 \ln(a/a_{*0})]} 
\label{expdelta-x}
\end{eqnarray}
%----------------------
%[e,l,d]
to determine the  universal scaling functions 
$s_{12}(x)$, $s_{22}(x)$, and $s_{11}(x)$.
The sign ambiguity in $s_{12}(x)$ is resolved by using 
continuity together with the analytic result for 
$s_{12}(0)$ given by Eq.~(\ref{s12:x->0}).

The numerical results for the modulus $|s_{ij}(x)|$ 
and the phase $\arg s_{ij}(x)$ of each of these functions 
are shown in Fig.~\ref{fig:sij-x}. 
The modulus $|s_{12}(x)|$ remains close to 1 for $x< 1$
but it decreases to 0.59 at $x=10$. 
The moduli $|s_{11}(x)|$ and $|s_{22}(x)|$ 
both have the tiny value 0.0018 at $x=0$
and they remain tiny for $x<1$.
While $|s_{22}(x)|$ increases to 0.37 at $x=10$,
$|s_{11}(x)|$ increases much more slowly to 0.035.
%{\bf [An increase of the cutoff $\Lambda_0$ has no
%effect on our results.]}
%{\bf [When the cutoff in our calculation is increased, the results
%for the $s_{ij}$ change by less than XX\%.]}

\subsection{Three-body recombination rates}

In this subsection, we use the universal scaling functions 
calculated in the previous sections to calculate the 3-body 
recombination rate into the shallow dimer.
We show the contributions from angular momentum $J=0$ through 6
as functions of the collision energy.

%%%%%%%%%%%%%%%%%%%%%%%%%%%%%%%%%%%%%%%%%%%%%%%%%%
\begin{figure}[htb]
\centerline{ \includegraphics*[width=10cm,angle=0,clip=true]{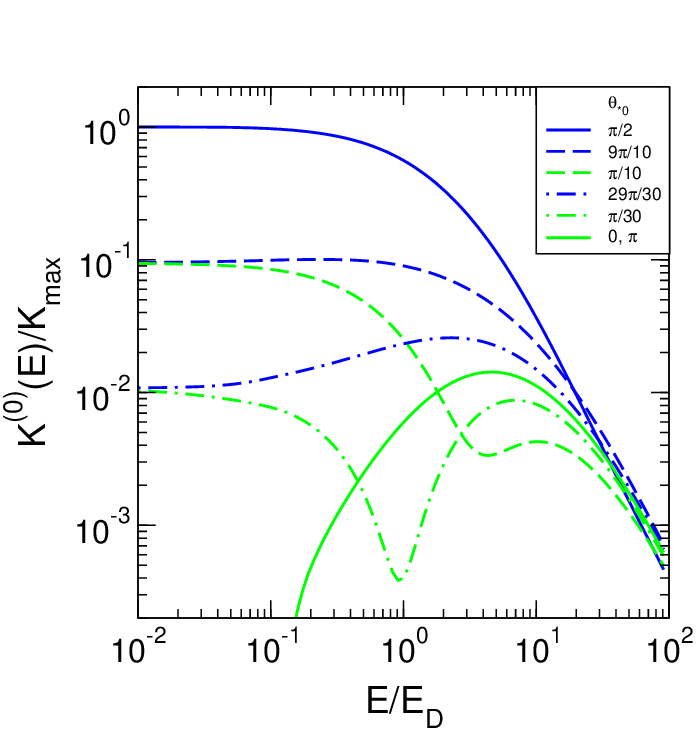}}
%\centerline{\includegraphics*[width=12cm,angle=0,clip=true]{data.eps}}
\vspace*{0.0cm}
\caption{(Color online) 
The $J=0$ contribution to the 3-body recombination rate into the 
shallow dimer for the case in which there are no deep dimers. 
The hyperangular average $K^{(0)}(E)$ is shown as a function of the 
collision energy $E$ for 6 values of $\theta_{*0}$:
$\frac{1}{2} \pi$ (upper solid line),
$\frac{9}{10} \pi$ and $\frac{1}{10} \pi$ (upper and lower dashed lines),
$\frac{29}{30} \pi$ and $\frac{1}{30} \pi$ (upper and lower dash-dotted
lines), 0 or $\pi$ (lower solid line).
}
\label{fig:alpha0}
\end{figure}
%%%%%%%%%%%%%%%%%%%%%%%%%%%%%%%%%%%%%%%%%%%%%%%%%%

In Eq.~(\ref{sigbr:rl}), the $J=0$ contribution $K^{(0)}(E)$ 
to the 3-body recombination rate is
expressed in terms of the universal scaling functions 
$s_{11}(x)$, $s_{12}(x)$, and $s_{22}(x)$.
In Fig.~\ref{fig:alpha0},
it is shown as a function of the collision energy $E$ 
for 6 values of $\theta_{*0}$: 0, $\frac{1}{30} \pi$, $\frac{1}{10} \pi$, 
$\frac{1}{2} \pi$, $\frac{9}{10} \pi$, and $\frac{29}{30} \pi$.
For $\theta_{*0} = \frac{1}{2} \pi$, $K^{(0)}(E)$ 
at $E=0$ has the maximum possible value $K_{\rm max}$
given in Eq.~(\ref{Kmax}) and it decreases monotonically with $E$.
For $\theta_{*0} = 0$ or $\pi$,
$K^{(0)}(E)$ vanishes at $E=0$, increases to a maximum value of 
$0.0143~K_{\rm max}$ at $E = 4.61~E_D$,
and then decreases.
For $\theta_{*0}$ between $\frac{1}{2} \pi$ and $\pi$,
$K^{(0)}(E)$ has its maximum at a nonzero energy.
As $\theta_{*0}$ increases from $\frac{1}{2} \pi$ to $\pi$,
the local minimum at $E=0$ becomes increasingly deep 
and the maximum moves outward from 0 to $4.61~E_D$.
For $\theta_{*0}$ between $0$ and $0.124 \pi$,
$K^{(0)}(E)$ has a local minimum at a nonzero energy $E$.
As $\theta_{*0}$ increases from $0$ to $0.124 \pi$,
the local minimum moves outward from 0 to about $8~E_D$
and its depth decreases until it becomes just an inflection point.
For $\theta_{*0}$ between $0.124 \pi$ and $\frac{1}{2} \pi$,
$K^{(0)}(E)$ decreases monotonically with $E$.

At small energies, the dimensionless recombination rate 
$K^{(0)}(E)/K_{\rm max}$
varies by many orders of magnitude as $a_{*0}$
ranges over a complete discrete scaling cycle.
Even at $E = E_D$, it varies by more than 3 orders of magnitude.
The variations with $a_{*0}$ become smaller at larger energy.
If there were no relevant length scales at high energy,
dimensional analysis would imply that  $K^{(0)}(E)$
should be proportional to $E^{-2}$ at large $E$.
Our results at the largest values of $E$ are compatible 
with the approach to this simple scaling behavior.
The differences between $K^{(0)}(E)$ for different values of $a_{*0}$ 
seem to decrease as a higher power of $E$.   
Our results at the largest values of $E$ are compatible 
with their approach to the asymptotic behavior $E^{-5/2}$.

%%%%%%%%%%%%%%%%%%%%%%%%%%%%%%%%%%%%%%%%%%%%%%%%%%
\begin{figure}[htb]
\centerline{ 
\includegraphics*[width=10cm,angle=0,clip=true]{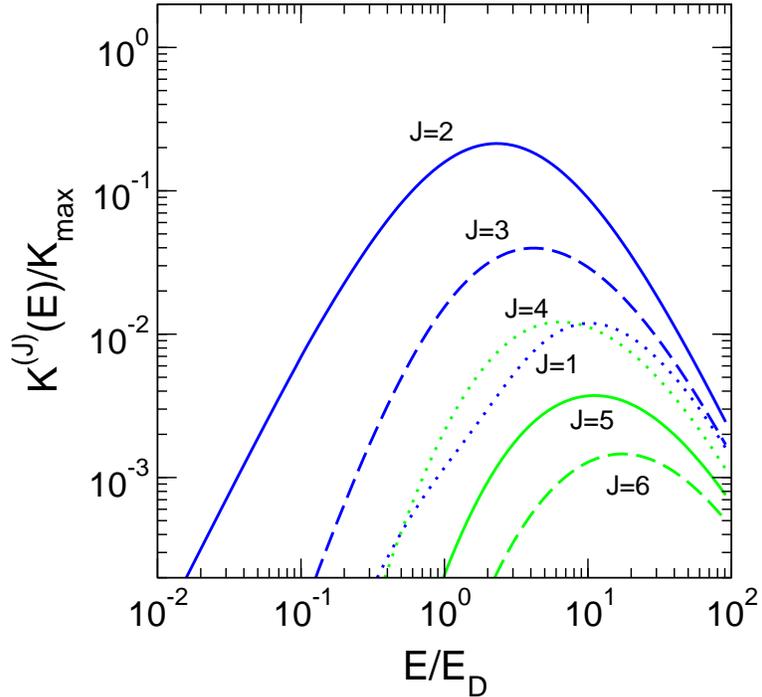}}
\vspace*{0.0cm}
\caption{(Color online) 
The contributions to the 3-body recombination rate into the 
shallow dimer from angular momenta $J=1,\ldots,6$. 
The hyperangular averages $K^{(J)}(E)$ are shown as a function of the 
collision energy $E$.}
\label{fig:alpha1-6}
\end{figure}
%%%%%%%%%%%%%%%%%%%%%%%%%%%%%%%%%%%%%%%%%%%%%%%%%%

In Eq.~(\ref{alpha-J:uni}), the contribution $K^{(J)}(E)$ 
to the 3-body recombination rate from angular momentum $J$ is
expressed in terms of the universal scaling functions $f_J(x)$.
In Fig.~\ref{fig:alpha1-6}, the rates $K^{(J)}(E)$ for $J$ 
from 1 to 6 are shown as functions of the collision energy $E$.
Over the range from 0 to 100~$E_D$, the largest of these 
contributions is the $J=2$  term.  It reaches its maximum value 
of $0.214~K_{\rm max}$ at $E = 4.18~E_D$.
The total contribution from $J=1$, 3, 4, 5, and 6 
is less than 10\% of the $J=2$ contribution if $E < 0.78~E_D$.
Thus the sum of the terms for $J=0$ and 2
is a good approximation to the recombination rate if $E < 0.78~E_D$.
The sum of the contributions from $J=5$ and 6
is less than 10\% of the sum of the contribution from $J=1$ through 4
if $E < 32~E_D$.  Thus the sum of the terms for $J=0$ through 4 
is a good approximation to the recombination rate if $E < 32~E_D$.
Truncations of the partial wave expansion quickly become
inaccurate at higher energies.

\section{Application to $\bm{^4}$He atoms}
\label{sec:helium}

The results from the previous section can be applied directly
to systems of $^4$He atoms.%
\footnote{A convenient conversion constant for $^4$He atoms is
$\hbar^2/m = 43.2788 \, {\rm K}~a_0^2 $.}
There are several modern potentials 
that are believed to describe the interactions of $^4$He atoms 
with low energy accurately.  
They all support a single weakly-bound diatomic molecule (or dimer).
The effective ranges for all these potentials are 
approximately $r_s\approx 14~a_0$, which is a little larger than
the van der Waals length scale $(m C_6/\hbar^2)^{1/4} = 10.2~a_0$.
The scattering lengths are larger by about a factor of ten, 
and they vary among the potentials.
The large scattering length, $a \gg r_s$, implies that $^4$He atoms
have universal properties that are determined by $a$
and the Efimov parameter $\kappa_*$.
First-order range corrections to the universal 
results are estimated to be suppressed by
roughly $r_s/a$ if $E \lesssim E_D$ and 
by  $r_s\,(mE/\hbar^2)^{1/2} = x r_s/a$ for $E \gtrsim E_D$.
This makes $^4$He atoms an ideal system to illustrate the universal
properties of atoms with large scattering length.
A thorough analysis of the universal properties of $^4$He atoms
has been presented by Braaten and Hammer \cite{BH02}.
Various 3-body observables for $^4$He atoms have been calculated by 
Platter and Phillips to next-to-next-to-leading order in the expansion 
in powers of $r_s/a$ \cite{Platter:2006ev}. 
The agreement with numerical results from exact solutions of the 3-body 
Schr\"odinger equation \cite{RY00,Ro03} is impressive.

\begin{figure}[htb]
\centerline{ \includegraphics*[width=15cm,angle=0,clip=true]{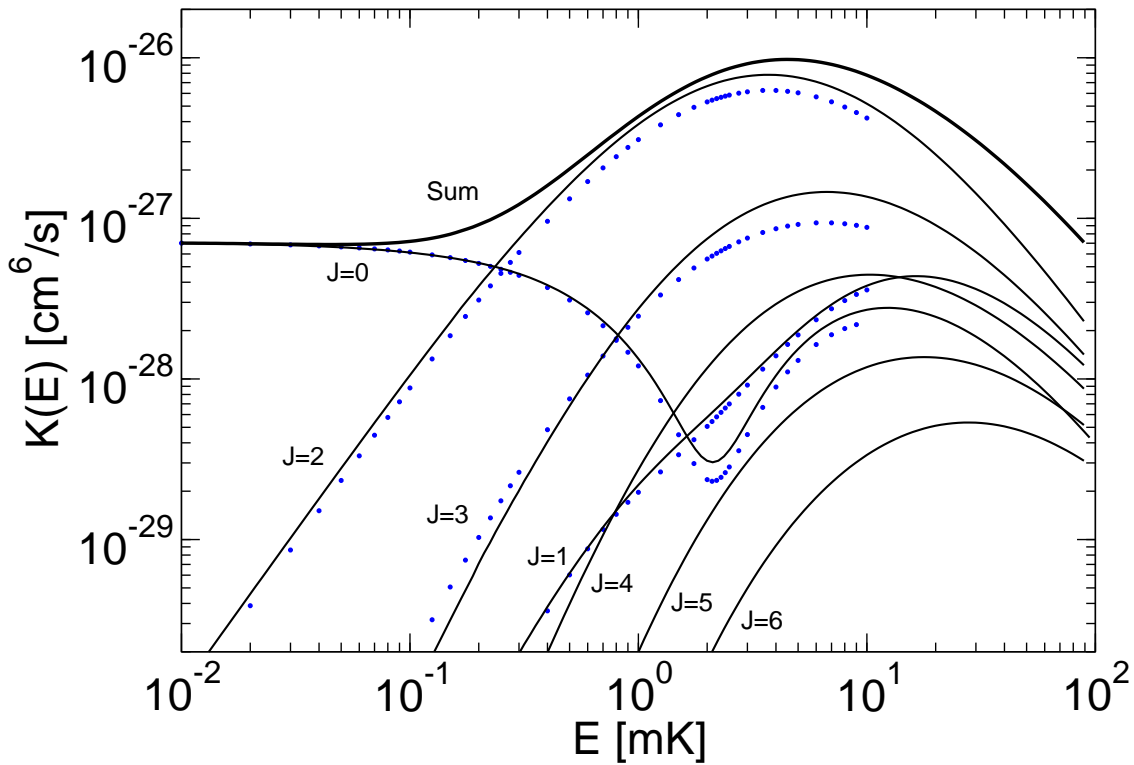}}
%\centerline{\includegraphics*[width=12cm,angle=0,clip=true]{data.eps}}
\vspace*{0.0cm}
\caption{(Color online) The 3-body recombination rates $K^{(J)}(E)$ 
for $^4$He atoms (in units of ${\rm cm}^6/{\rm s}$) 
as functions of the collision energy $E$ (in units of mK) for $J=0$ through 6.
The solid lines are our zero-range results for $a=164.5~a_0$ 
and $a_{*0} = 143.1~a_0$.
The dots are the results obtained in Ref.~\cite{SEGB02} 
by solving the 3-body Schr\"odinger equation for the HFD-B3-FCI1 potential.}
\label{fig:alpha-He}
\end{figure}

The 3-body recombination rate for $^4$He atoms was first calculated 
by Suno, Esry, Greene, and Burke \cite{SEGB02}.
They solved the 3-body Schr\"odinger equation for $^4$He atoms 
interacting through the HFD-B3-FCI1 potential \cite{AJM95}.
They calculated $K_{\rm shallow}(E)$ as a function of the collision energy
from the threshold to 10 mK \cite{SEGB02},
including the contributions with angular momentum quantum number 
$J=0$, 1, 2, and 3.
Their results are shown as dots in Fig.~\ref{fig:alpha-He}.
Shepard has calculated the 3-body recombination rates of $^4$He atoms 
for the HFD-B3-FCI1 and other $^4$He potentials using separable potentials
that reproduce the corresponding two-body phase shifts
together with a 3-body force that is adjusted to fit the 
binding energy of the excited trimer \cite{Shepard:2007gj}.

%%%%%%%%%%%%%%%%%%%%%%%%%%%%%%%%%%%%%%%%%%%%%%%%%%
%%%%%%%%%%%%%%%%%%%%%%%%%%%%%%%%%%%%%%%%%%%%%%%%%%

To apply the zero-range predictions for the 3-body recombination rate 
to $^4$He atoms interacting through the HFD-B3-FCI1 potential, 
we must determine the scattering length $a$ and the 3-body parameter
$a_{*0}$ for that potential.
The scattering length for the HFD-B3-FCI1 potential
is $a = 172~a_0$ and the binding energy of the shallow dimer 
is $E_D = 1.600$ mK \cite{private}.  
The zero-range predictions for low-energy 3-body observables are most accurate
if the dimer bound state pole is at the correct position \cite{BH02}. 
Therefore, we use the  dimer binding energy 
as the 2-body input instead of $a$.
Using the zero-range expression for the dimer binding energy 
in Eq.~(\ref{E-dimer}), we obtain the scattering length
%----------------------
\begin{equation}
a^{\rm He} = 164.5~a_0 \,.
\label{eq:aHe}
\end{equation}
%----------------------
%[hw,d,l]
We can determine the value of $a_{*0}$ for $^4$He atoms 
interacting through the HFD-B3-FCI1 potential from the value 
of the 3-body recombination loss rate at threshold:  
$K_{\rm shallow}(0) =  7.10 \times 10^{-28}$ cm$^6$/s.  Inserting 
$a^{\rm He} = 164.5~a_0$ in Eq.~(\ref{K0-analytic})
and solving for $a_{*0}$, we obtain 
%----------------------
\begin{equation}
a^{\rm He}_{*0} = 143.1~a_0.
\label{astar0He}
\end{equation}
%----------------------
%[hw,d,l]
The result for $a^{\rm He}_{*0}$ is approximately equal to 
$0.870 \,a^{\rm He}$.
The near equality between $a^{\rm He}_{*0}$ and $a^{\rm He}$
reflects the fact that $K_{\rm shallow}(0)$ for $^4$He is much smaller 
than the maximum value 
$6 C_{\rm max} \hbar (a^{\rm He})^4/m =  3.67 \times 10^{-26}$ cm$^6$/s
allowed by universality, which implies that 
$a/a_{*0}$ is close to a zero of the sine function in 
the numerator of Eq.~(\ref{K0-analytic}).
The value of $a_{*0}$ for $^4$He can also be determined 
from the binding energy of the excited $^4$He trimer, which is
$E_3^{(1)} = 2.62$ mK for the HFD-B3-FCI1 potential \cite{private}.
The resulting value, $a_{*0} \approx 146~a_0$, is consistent 
with the value in Eq.~(\ref{astar0He}).  Its accuracy is limited 
by the accuracy to which the numerical value of the constant 
in Eq.~(\ref{a*0-kappa*}) is known.

The zero-range predictions for $K^{(J)}(E)$ are given in 
Eq.~(\ref{sigbr:rl}) for $J=0$ and in Eq.~(\ref{alpha-J:uni}) for $J \ge 1$.
The only information about the HFD-B3-FCI1 potential that we use
are the values of $a$ and $a_{*0}$ given in 
Eqs.~(\ref{eq:aHe}) and (\ref{astar0He}).
Our results for $K^{(J)}(E)$, $J=0,1,\ldots,6$, are shown 
as solid lines in Fig.~\ref{fig:alpha-He}.
The $J=0$ term dominates at low energies, but has a dramatic dip near 
2 mK, where it is smaller than its value at $E = 0$ by about a factor of 30.
The $J=2$ term takes over as the dominant term when the 
energy exceeds about 0.2 mK.  When it reaches its maximum value 
at an energy near 4 mK, other partial waves begin to give important
contributions.

Our results can be compared with those of Ref.~\cite{SEGB02} 
for $J=0$, 1, 2, and 3, which are shown as dots in Fig.~\ref{fig:alpha-He}.
The qualitative agreement is excellent.
The deep local minimum for $J=0$ and the peculiar shape for $J=1$
are both reproduced by the zero-range results. 
There is also good quantitative agreement for $J=0$, 1, and 2.
At $E = E_D = 1.6$ mK, 
the fractional discrepancies for $J=0$, 1, and 2 are 18.5\%, 14.4\%,
and 20.3\%, respectively.
This is a little larger than the expected fractional error from range 
corrections, which is $r_s/a = 8.5\%$. 
At $E = 10$ mK, which is the highest energy for which the
3-body recombination rate was calculated in Ref.~\cite{SEGB02}, 
the fractional discrepancies for $J=0$, 1, and 2 are
16.5\%, 6.0\%, and 18.6\%.  They are consistent with the expected 
fractional error from range corrections, which is 
$xr_s/a = 21\%$.  For $J=3$, 
the agreement between our results and those of 
Ref.~\cite{SEGB02} is not as good.  The fractional discrepancy 
varies from $+77\%$ at 0.1 mK to $-36\%$ at 10 mK.
Shepard has also calculated the contributions to the 3-body 
recombination rate for the HFD-B3-FCI1 potential 
from partial waves up to $J=3$ \cite{Shepard:2007gj}.
His results for $J=3$ are similar to ours and show the same
discrepancy with Ref.~\cite{SEGB02}.  The agreement between 
our results for $J=3$ and those of Ref.~\cite{Shepard:2007gj}
gives us confidence in the accuracy of our numerical calculations.
We do not understand the origin of the discrepancy
with the results of Ref.~\cite{SEGB02}.

The inclusion of higher order corrections in $r_s/a$ should 
improve the agreement with the results of Ref.~\cite{SEGB02}
at all energies. If the first order corrections in $r_s/a$ are
included, the errors should decrease to roughly $(r_s/a)^2 = 0.7\%$
for $E \lesssim E_D= 1.6$ mK and to roughly $(xr_s/a)^2$ at higher energies.
A first calculation of the range corrections
to the threshold recombination rate $K_{\rm shallow}(0)$ was  
carried out in Ref.~\cite{Hammer:2006zs}.

\section{Effects of deep dimers}
\label{sec:deep}

In this section and in the subsequent section, 
we consider atoms that have deep dimers.
In the scaling (or zero-range) limit, 
the cumulative effect of all the deep dimers on Efimov physics 
can be taken into account through one
additional parameter $\eta_*$ \cite{Braaten:2003yc}. 
The atom-dimer phase shift and the hyperangular average 
of the 3-body recombination rate are determined by the same 
universal scaling functions as for $\eta_*=0$.

\subsection{Generalization of Efimov's Radial Law}

The existence of deep dimers implies that there are states in the 
3-atom sector with energies near the 3-atom threshold
that consist of an atom and a deep dimer with large kinetic energies.
We will denote these states by the symbol $\bmAD$.
These states provide inelastic atom-dimer scattering channels 
and additional 3-body recombination channels.
In the scaling limit, the only contributions to the recombination rate
into deep dimers are from the $J=0$ channels.
The contributions from $J\ge 1$ are suppressed by powers of $r_s/a$.
The hyperangular average of the inclusive 3-body recombination rate
into deep dimers can be expressed as
%----------------------
\begin{equation}
K_{\rm deep}(E) = \frac{144 \sqrt{3} \pi^2\hbar^5}{m^3 E^2}
\sum_{n=3}^\infty \sum_{\bmAD}
\left| S_{AAA,\bmAD}^{(J=0,n)}(E) \right|^2,
\label{Kdeep-S}
\end{equation}
%----------------------
%[e,hw,l,d]
where $S_{AAA,\bmAD}^{(J=0,n)}(E)$ is the S-matrix element for the 
transition from a 3-atom scattering state labelled by $(J=0,n)$
to a specific state $A {\bm D}$ consisting of an atom and deep dimer.
The hyperangular average is implemented in Eq.~(\ref{Kdeep-S})
by the sum over $n$.

The states $\bmAD$ also have an effect on 
the amplitudes for atom-dimer elastic scattering and for
3-body recombination into the shallow dimer.  
These effects are strongly constrained by the fact 
that the states $\bmAD$ can be accessed only 
through a single adiabatic hyperspherical channel: 
the $J=0$ channel whose potential is attractive in the scale-invariant 
region of the hyperradius $R$.  If there are deep dimers,
an incoming hyperradial wave need not be totally reflected from
the short-distance region of $R$, because it can ultimately emerge 
as a scattering state consisting of an atom and a deep dimer.
The reflected hyperradial wave will therefore differ from the incident 
hyperradial wave not only by a phase shift 
$e^{2 i \theta_{*0}}$ but also by a decrease 
in its amplitude by a factor $e^{- 2 \eta_*}$.
The parameter $\eta_*$, which was introduced in 
Ref.~\cite{Braaten:2002sr}, determines the widths of Efimov trimers 
as well as all other effects of the deep dimers on low-energy
3-atom scattering processes.  If the zero-range result 
for an amplitude for the case of no deep dimers 
is known as an analytic function of $a_{*0}$, 
the corresponding amplitude for a system with deep
dimers can be obtained without any additional calculation simply
by making the substitution   
%----------------------
\begin{eqnarray}
\ln a_{*0} \longrightarrow \ln a_{*0} - i \eta_*/s_0 \,.
\label{logsub}
\end{eqnarray}
%----------------------
%[hw,e,d,l]
The generalization of Efimov's Radial Laws in Eqs.~(\ref{RL})
to the case of atoms with deep dimers are
%----------------------
\begin{subequations}
\begin{eqnarray}
S_{AD,AD}^{(J=0)}(E) & = & s_{22}(x)
+  \frac{s_{21}(x)^2 e^{2i \theta_{*0} -2 \eta_*}} 
        {1 - s_{11}(x) e^{2i \theta_{*0} -2 \eta_*}} \,,
\label{RLdeep:AD}
\\
S_{AD,AAA}^{(J=0,n)}(E) & = & s_{2n}(x) 
+  \frac{s_{21}(x) s_{1n}(x) e^{2i \theta_{*0} -2 \eta_*}} 
       {1 - s_{11}(x) e^{2i \theta_{*0} -2 \eta_*}} \,.
\label{RLdeep:ADAAA}
\end{eqnarray}
\label{RLdeep}
\end{subequations}
%----------------------
%[hw,e,l,d]
Thus these S-matrix elements are determined by the same
universal scaling functions $s_{ij}(x)$ 
as in the case $\eta_* = 0$ with no deep dimers. 

The S-matrix elements for transitions from an atom-dimer scattering 
state or from a 3-atom scattering state into a specific state
$\bmAD$ consisting of an atom and a deep dimer
depend on details of physics at short distances.
However as pointed out in Refs.~\cite{Braaten:2002sr,Braaten:2003yc}, 
the inclusive probability 
summed over all such states $\bmAD$ is insensitive to short distances 
and is determined completely by the parameters $a$, 
$\kappa_*$ and $\eta_*$.  The inclusive probabilities for 
producing an atom and a deep dimer from an atom-dimer
collision or from the collision of three atoms 
in the state labelled $(J=0,n)$ are
%----------------------
\begin{subequations}
\begin{eqnarray}
\sum_{\bmAD} | S_{AD,\bmAD} ^{(J=0)}(E)|^2 &=& 
\frac{(1 - e^{-4 \eta_*}) |s_{21}(x)|^2} 
    {| 1 - s_{11}(x)  e^{2i \theta_{*0} -2 \eta_*} |^2} \,,
\label{sum:SADX}
\\
\sum_{\bmAD} | S_{AAA,\bmAD}^{(J=0,n)} (E)|^2 &=& 
\frac{(1 - e^{-4 \eta_*}) |s_{n1}(x)|^2} 
    {|1 - s_{11}(x) e^{2i \theta_{*0} -2 \eta_*} |^2} \,.
\label{sum:SAAAX}
\end{eqnarray}
\label{sum:SX}
\end{subequations}
%----------------------
%[hw,e,l,d]
The factor $s_{21}$ in Eq.~(\ref{sum:SADX}) is the amplitude 
for an incoming atom-dimer scattering state 
to be transmitted through the scaling region into a 
hyperradial wave in the scale-invariant region.
The factor $1/(1 - s_{11} e^{2i \theta_{*0} -2 \eta_*})$
takes into account an arbitrary number of reflections 
of the hyperradial wave from 
the short-distance region and then from the scaling region.
The factor $1 - e^{-4 \eta_*}$ is the probability that a hyperradial wave
incident on the short-distance region will not be reflected 
and will therefore ultimately emerge as a scattering state 
of an atom and a deep dimer.  The analog of the unitarity
condition in Eq.~(\ref{unitarity:0}) for the case $\eta_* > 0$ is 
%----------------------
\begin{equation}
|S_{AD,AD}^{(J=0)}(E)|^2 
+ \sum_{n=3}^\infty |S_{AD,AAA}^{(J=0,n)}(E)|^2 
+ \sum_{\bmAD} |S_{AD,\bmAD}^{(J=0)}(E)|^2 = 1 .
\label{unitarity:eta*}
\end{equation}
%----------------------
%[e,hw,l,d]
If we insert the S-matrix elements in Eqs.~(\ref{RLdeep})
and the probability in Eq.~(\ref{sum:SADX}),
we can use the unitarity conditions for the matrix $s_{ij}$ 
in Eq.~(\ref{s:unitary})
to show that Eq.~(\ref{unitarity:eta*}) is automatically satisfied.

The hyperangular average
of the 3-body recombination rate into the shallow dimer
is obtained by squaring the S-matrix element in Eq.~(\ref{RLdeep:ADAAA}),
summing over the index $n$ labelling the 3-atom states,
and then multiplying by a kinematic factor.
Using the unitarity condition in Eq.~(\ref{unitarity:eta*}),
the expression for the S-matrix element in Eq.~(\ref{RLdeep:AD}),
and the expression for the probability in Eq.~(\ref{sum:SADX}),
this can be expressed as
%----------------------
\begin{equation}
K^{(0)}(E) = \frac{144 \sqrt{3} \pi^2}{x^4}
\left( 1 
- \left| s_{22}(x) + \frac{s_{12}(x)^2 e^{2i \theta_{*0} -2 \eta_*}}
                     {1 - s_{11}(x) e^{2i \theta_{*0} -2 \eta_*}} \right|^2
- \frac{(1 - e^{-4 \eta_*}) |s_{12}(x)|^2}
      {|1 - s_{11}(x) e^{2i \theta_{*0} -2 \eta_*}|^2} \right) 
\frac{\hbar a^4}{m} \,.
\label{Kshallow-s}
\end{equation}
%----------------------
%[e,hw,l,d]
This reduces to Eq.~(\ref{sigbr:rl}) in the limit $\eta_* \to 0$.
The hyperangular average
of the inclusive 3-atom recombination rate into deep dimers
is given by summing the probability in Eq.~(\ref{sum:SAAAX})
over the index $n$ labelling the 3-atom states
and then multiplying by a kinematic factor.
Using the unitarity of the matrix $s_{ij}(x)$,
the rate can be expressed as
%----------------------
\begin{eqnarray}
K_{\rm deep}(E) = 
\frac{144 \sqrt{3} \pi^2 (1 - e^{-4 \eta_*}) 
	\big( 1 - |s_{11}(x)|^2 - |s_{12}(x)|^2 \big)} 
    {x^4 |1 - s_{11}(x) e^{2i \theta_{*0} -2 \eta_*} |^2} \, 
\frac{\hbar a^4}{m} \,.
\label{Kdeep-s}
\end{eqnarray}
%----------------------
%[e,hw,l,d]
This contribution vanishes in the limit $\eta_* \to 0$.
The recombination rates in Eqs.~(\ref{Kshallow-s})
and (\ref{Kdeep-s})
are completely determined by the same universal scaling functions
$s_{11}(x)$, $s_{12}(x)$, and $s_{22}(x)$ 
that determine the 3-body recombination rate into 
shallow dimers in the case $\eta_*=0$.

\subsection{Analytic results at threshold}

Analytic expressions for the 3-body recombination rates 
$K_{\rm shallow}(0) = K^{(0)}(0)$ and $K_{\rm deep}(0)$ at the 
threshold $E=0$ can be obtained by inserting the analytic results 
for the universal scaling functions $s_{11}(x)$, $s_{12}(x)$, 
and $s_{22}(x)$ in Eqs.~(\ref{sij:x->0}) into Eqs.~(\ref{Kshallow-s})
and (\ref{Kdeep-s}).
The 3-body recombination rates at threshold into the shallow dimer 
and into deep dimers are
%----------------------
\begin{subequations}
\begin{eqnarray}
K_{\rm shallow}(0) &=& 
\frac{768 \pi^2 (4 \pi - 3 \sqrt{3})
(\sin^2 [s_0 \ln(a/a_{*0})]  + \sinh^2\eta_*)}
{\sinh^2(\pi s_0 + \eta_*) + \cos^2 [s_0 \ln(a/a_{*0})]}\,
\frac{\hbar a^4}{m} \,,
\label{K-shallow:eta}
\\
K_{\rm deep}(0) &=& 
\frac{384 \pi^2 (4 \pi - 3 \sqrt{3}) \coth(\pi s_0) \sinh(2\eta_*)}
    {\sinh^2(\pi s_0 + \eta_*) + \cos^2 [s_0 \ln(a/a_{*0})]}\,
\frac{\hbar a^4}{m} \,.
\label{K-deep:eta}
\end{eqnarray}
\label{K:eta}
\end{subequations}
%----------------------
%[d,hw,l,e]
Since $\sinh^2(\pi s_0+ \eta_*) > 139$ is so large,
these expressions can be approximated 
with errors of less than 1\% by omitting the $\cos^2$ terms 
in the denominators.  To within the same 
accuracy, we can approximate $\sinh(\pi s_0 + \eta_*)$,
$\sinh(\pi s_0)$, and $\cosh(\pi s_0)$
by exponentials to get the simpler expressions
%----------------------
\begin{subequations}
\begin{eqnarray}
K_{\rm shallow}(0) &\approx&
6 C_{\rm max} e^{-2 \eta_*}
\left( \sin^2 [s_0 \ln(a/a_{*0})]  + \sinh^2\eta_* \right) \, \hbar a^4/m ,
\label{K-shallow:approx}
\\
K_{\rm deep}(0) &\approx& 
\mbox{$\frac{3}{2}$} C_{\rm max} (1 - e^{-4\eta_*})  \, \hbar a^4/m,
\label{K-deep:approx}
\end{eqnarray}
\label{K:approx}
\end{subequations}
%----------------------
%[d,hw,l,e]
where $C_{\rm max} \approx 67.1$ is given in Eq.~(\ref{K-max}).

\subsection{Three-body recombination rate}

In this subsection, we use the universal scaling functions 
calculated in the previous section to calculate the 3-body 
recombination rate into deep dimers and the $J=0$ contribution 
to the 3-body recombination rate into the shallow dimer.
The effects of deep dimers on the contributions to the 3-body 
recombination rate into the 
shallow dimer from $J\ge 1$ are suppressed in the zero-range limit.

%%%%%%%%%%%%%%%%%%%%%%%%%%%%%%%%%%%%%%%%%%%%%%%%%%
\begin{figure}[htb]
\includegraphics*[width=7.5cm,angle=0,clip=true]{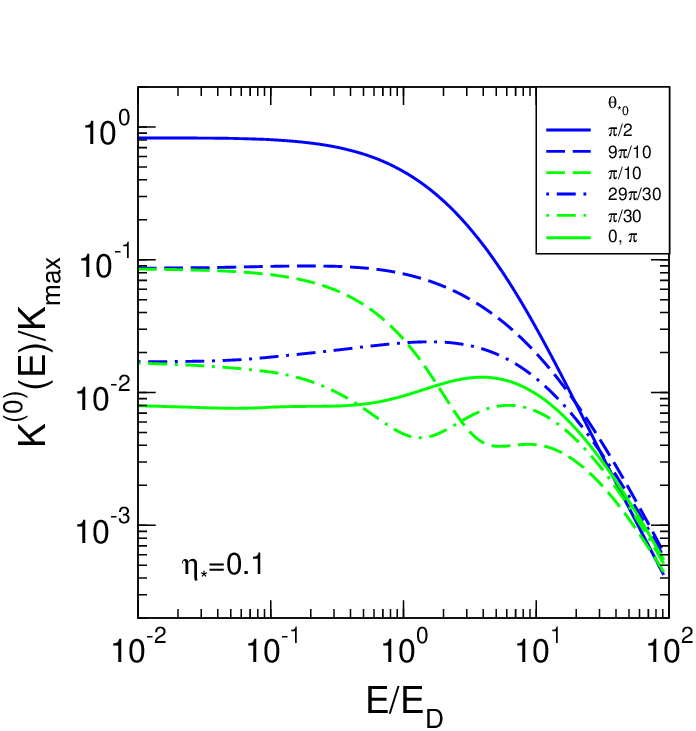}
\hspace{1cm}
\includegraphics*[width=7.5cm,angle=0,clip=true]{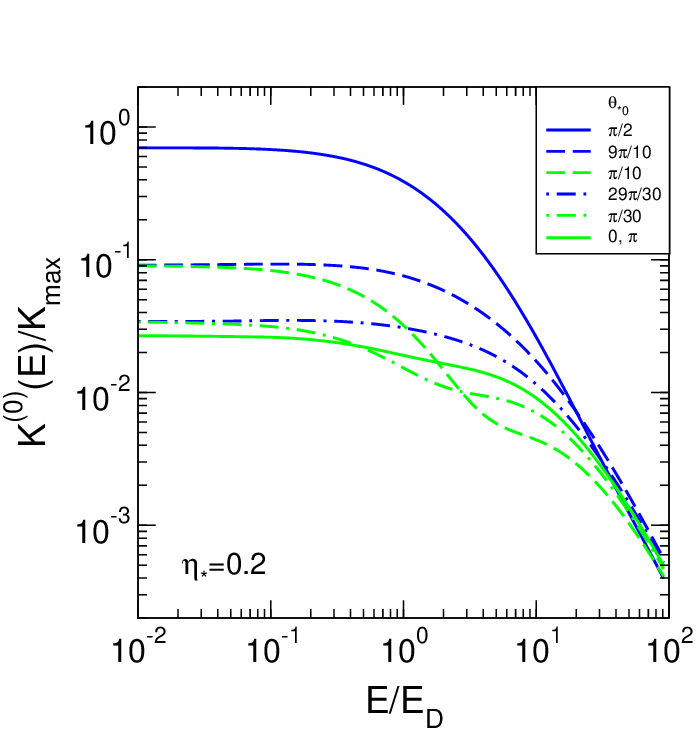}
\vspace*{0.0cm}
\caption{(Color online) 
The $J=0$ contribution to the 3-body recombination rate into the 
shallow dimer for cases in which there are one or more deep dimers. 
The hyperangular average $K^{(0)}(E)$ is shown as a function of the 
collision energy $E$ or 6 values of $\theta_{*0}$:
$\frac{1}{2} \pi$ (upper solid line),
$\frac{9}{10} \pi$ and $\frac{1}{10} \pi$ (upper and lower dashed lines),
$\frac{29}{30} \pi$ and $\frac{1}{30} \pi$ (upper and lower dash-dotted
lines), 0 or $\pi$ (lower solid line).
The left and right panels are for $\eta_* = 0.1$ and $\eta_* = 0.2$,
respectively.}
\label{fig:alpha0eta}
\end{figure}
%%%%%%%%%%%%%%%%%%%%%%%%%%%%%%%%%%%%%%%%%%%%%%%%%%

In Eq.~(\ref{Kshallow-s}), the $J=0$ contribution $K^{(0)}(E)$ 
to the 3-body recombination rate into the shallow dimer is
expressed in terms of the universal scaling functions 
$s_{11}(x)$, $s_{12}(x)$, and $s_{22}(x)$.
In Fig.~\ref{fig:alpha0eta},
it is shown as a function of the collision energy $E$
for 6 values of $\theta_{*0}$: 0, $\frac{1}{30} \pi$, $\frac{1}{10} \pi$, 
$\frac{1}{2} \pi$, $\frac{9}{10} \pi$, and $\frac{29}{30} \pi$.
The left and right panels are for $\eta_{*0} = 0.1$ and 0.2, 
respectively.
Comparing with Fig.~\ref{fig:alpha0} for $\eta_{*0} = 0$,
we see that the local minimum for $\theta_{*0}$ between 0 and $0.124~a_0$
becomes less pronounced as $\eta_*$ 
increases.  When $\eta_*$ is large enough, 
the local minimum disappears altogether
and $K^{(0)}(E)$ becomes a monotonically 
decreasing function of $E$.

%%%%%%%%%%%%%%%%%%%%%%%%%%%%%%%%%%%%%%%%%%%%%%%%%%
\begin{figure}[htb]
\centerline{ \includegraphics*[width=10cm,angle=0,clip=true]{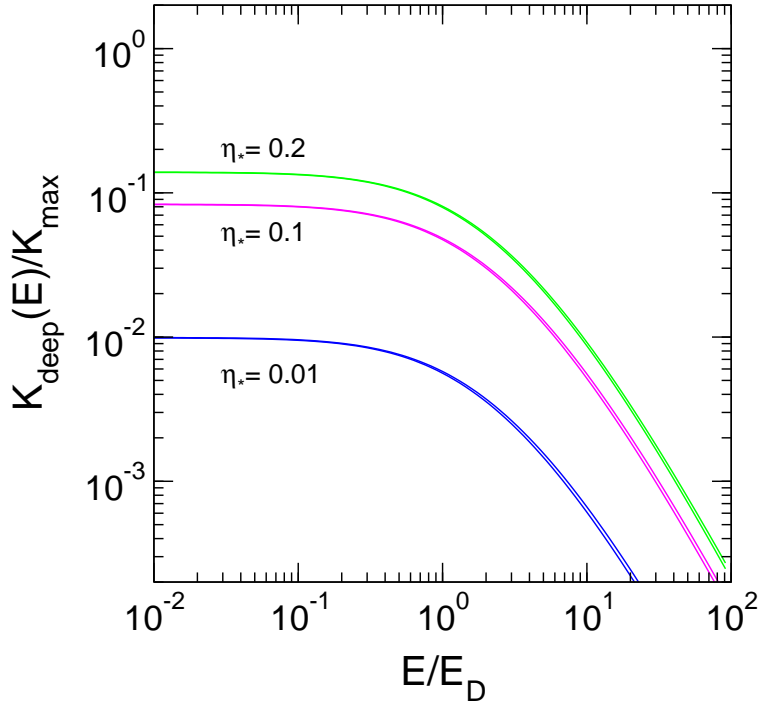}}
%\centerline{\includegraphics*[width=12cm,angle=0,clip=true]{data.eps}}
\vspace*{0.0cm}
\caption{(Color online) 
Three-body recombination rate into deep dimers. 
The maximum and minimum values of $K_{\rm deep}(E)$ 
with respect to variations of $\theta_{*0}$ are shown as functions of the 
collision energy $E$ for three values of $\eta_*$: 0.01, 0.1, and 0.2.}
\label{fig:alphadeep}
\end{figure}
%%%%%%%%%%%%%%%%%%%%%%%%%%%%%%%%%%%%%%%%%%%%%%%%%%

In Eq.~(\ref{Kdeep-s}), the 3-body recombination rate into
deep dimers is expressed in terms of the universal scaling functions 
$s_{11}(x)$ and $s_{12}(x)$.
In Fig.~\ref{fig:alphadeep}, the maximum and minimum values of 
$K_{\rm deep}(E)$ with respect to variations of $\theta_{*0}$ are 
shown for three values of $\eta_{*0}$: 0.01, 0.1, and 0.2.
The differences between the maximum and minimum values 
are only visible at energies greater than $E_D$.  
The 3-body recombination rate into
deep dimers is insensitive to the value of $\theta_{*0}$,
because the dependence on $\theta_{*0}$ in Eq.~(\ref{Kdeep-s})
is suppressed by a 
factor of $s_{11}(x)$, which remains small for $x < 10$.
For any value of $\eta_*$, $K_{\rm deep}(E)$
decreases monotonically with $E$.  
If there were no relevant length scales at high energy,
dimensional analysis would imply that $K_{\rm deep}(E)$
should be proportional to $E^{-2}$ at large $E$.
Our results at the largest values of $E$ are compatible 
with the approach to this simple scaling behavior.

\section{Application to $\bm{^{133}}$Cs Atoms}

The Innsbruck group has carried out measurements of the 
3-body recombination rate for ultracold $^{133}$Cs atoms%
\footnote{A convenient conversion constant for $^{133}$Cs atoms
is $\hbar^2/m = 1.30339 \, {\rm K}~a_0^2$.}
in the $|f=3, m_f=+3 \rangle$ hyperfine state \cite{Grimm06}.
By varying the magnetic field from 0 to 150 G, 
they were able to change the
scattering length from $-2500~a_0$ through 0 to $+1600~a_0$.
In this range of magnetic field, the $|f=3, m_f=+3 \rangle$ state
is the lowest hyperfine state, so 2-body losses are energetically 
forbidden.  Thus, the dominant loss mechanism is 3-body recombination.
The van der Waals length scale for Cs atoms
is $(mC_6/\hbar^2)^{1/4} \approx 200a_0$.
The range of scattering lengths studied by the Innsbruck group 
includes a universal region of large negative $a$ 
and a universal region of large 
positive $a$ separated by a nonuniversal region of small $|a|$.
In the two regions of large scattering length, few-body physics 
should be universal and characterized by 3-body parameters 
$\kappa_*$ and $\eta_*$ that may be different in 
each universal region.
An interesting open question is whether there is any relation
between the 3-body parameters $\kappa_*$ and $\eta_*$ for
different universal regions.

In the region of negative $a$, the Innsbruck group measured the 
loss rate constant $L_3$ as a function of $a$ at three different 
temperatures: $T=10$ nK, 200 nK, and 250 nK.
They observed a dramatic enhancement of the loss rate for $a$ 
near $-850~a_0$.  At $T = 10$ nK, 
the loss rate as a function of $a$ can be fitted rather well by the 
zero-range formula for $T=0$ in Ref.~\cite{Braaten:2003yc} 
with parameters $a_*' = -850(20)~a_0$ and $\eta_* = 0.06(1)$,
where $a_*'$ denotes the negative scattering length for which 
there is an Efimov resonance at the 3-atom threshold.
Thus, the large enhancement in the loss rate can be explained by the 
resonant enhancement from an Efimov trimer near the 3-atom threshold.
More recently, the Innsbruck group has measured the position of the
maximum loss rate as a function of the temperature \cite{grimm1107}.
Its behavior as a function of temperature can be explained at least 
qualitatively by the dependence of the binding energy and width 
of the Efimov resonance on the scattering length \cite{Jonsell06,YFT06}.

\begin{figure}[htb]
\centerline{\includegraphics*[width=12cm,angle=0,clip=true]{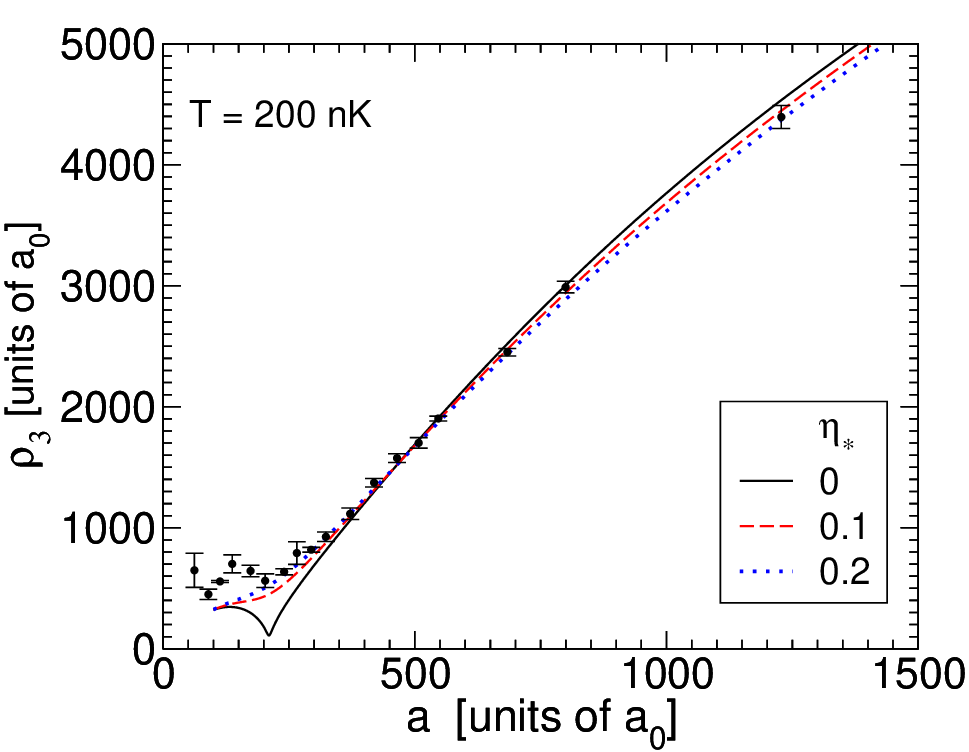}}
\vspace*{0.0cm}
\caption{(Color online) The 3-body recombination length $\rho_3$ 
for $^{133}$Cs atoms as a function of $a$ for $T=200$ nK.
The data points are from Ref.~\cite{Grimm06}.
The curves are the zero-range predictions for three values 
of $\eta_*$: 
0 (solid line), 0.1 (dashed line), and 0.2 (dotted line).}
\label{fig:alpha-Cs}
\end{figure}

In the region of positive $a$, the Innsbruck group measured the 
loss rate constant $L_3$ at $T=200$ nK for values of $a$ 
ranging from $62.3~a_0$ to $1228~a_0$.
Their results in this region are shown in Fig.~\ref{fig:alpha-Cs}.
The vertical axis is the recombination length $\rho_3$ defined by
\begin{equation}
\rho_3 = \left( \frac{2m}{\sqrt{3} \hbar} L_3 \right)^{1/4}.
\label{rho3}
\end{equation}
%
%[d,l,hw]
They observed a local minimum in the loss rate for $a$ 
near $200~a_0$.  This value is near the 
van der Waals length scale $(mC_6/\hbar^2)^{1/4} \approx 200~a_0$,
so range corrections may be large near the minimum.
The zero-range prediction for the loss rate at $T=0$ 
is $\alpha = K(0)/6$, where
$K(0)$ is the sum of $K_{\rm shallow}(0)$ and $K_{\rm deep}(0)$
in Eqs.~(\ref{K:eta}).  To an approximation of better than 1\%, 
they can be approximated by the expressions in Eqs.~(\ref{K:approx}).
By fitting the data for $a > 500~a_0$ to this
expression, they obtained $a_+ = 1060(70)~a_0$ for the 
scattering length $e^{\pi/(2 s_0)} a_{*0}$
at which the coefficient of $a^4$ achieves its maximum value.  
Universality would then imply that the minimum 
should be at $a_{*0} = 223(15)~a_0$.
The fit was insensitive to the value of $\eta_*$ and yielded 
only the upper bound $\eta_* < 0.2$.
The Innsbruck group also determined the location of the minimum directly 
by measuring the fraction of atoms that were lost after a fixed time.
The result was $a_{\rm min} = 210 (10)~a_0$, which is consistent 
with the value obtained by fitting the data for $a > 500~a_0$.

We now consider whether our results 
are applicable to this system of $^{133}$Cs atoms.
In the experiments with positive scattering length, the typical 
peak number density was $5 \times 10^{13}~{\rm cm}^{-3}$.
The corresponding critical temperature for Bose-Einstein condensation
is $T_c \approx 160$~nK.  Thus, the temperature $T = 200$~nK
was not far above $T_c$.  We ignore this complication 
and calculate thermal averages using the Maxwell-Boltzmann distribution 
as in Eq.~(\ref{alpha-T})  instead of the Bose-Einstein distribution.
When $a = a_{\rm min}$, the zero-range prediction
for the binding energy $E_D$ of the shallow dimer in Eq.~(\ref{E-dimer})
gives $3 \times 10^4$~nK, so $T /E_D \approx 0.007$.  
The binding energy $E_D$ decreases to about
$9 \times 10^2$~nK at $a = 1228~a_0$, so $T /E_D \approx 0.22$.  
Thus the temperature $T = 200$~nK 
is safely in the region $T < 30~E_D$ in which the 
thermal average in Eq.~(\ref{alpha-T}) can be calculated accurately
using the scaling functions calculated in Section~\ref{sec:unisf}.

%%%%%%%%%%%%%%%%%%%%%%%%%%%%%%%%%%%%%%%%%%%%%%%%%%
%%%%%%%%%%%%%%%%%%%%%%%%%%%%%%%%%%%%%%%%%%%%%%%%%%

\begin{figure}[htb]
\includegraphics*[width=7.5cm,angle=0,clip=true]{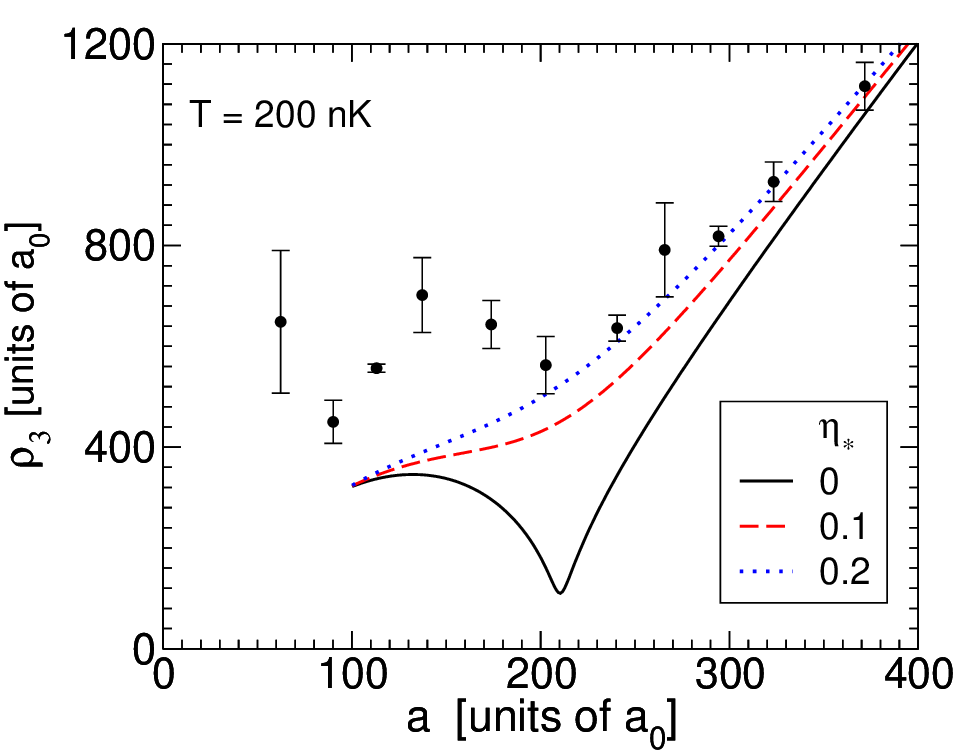}
\hspace{1cm}
\includegraphics*[width=7.5cm,angle=0,clip=true]{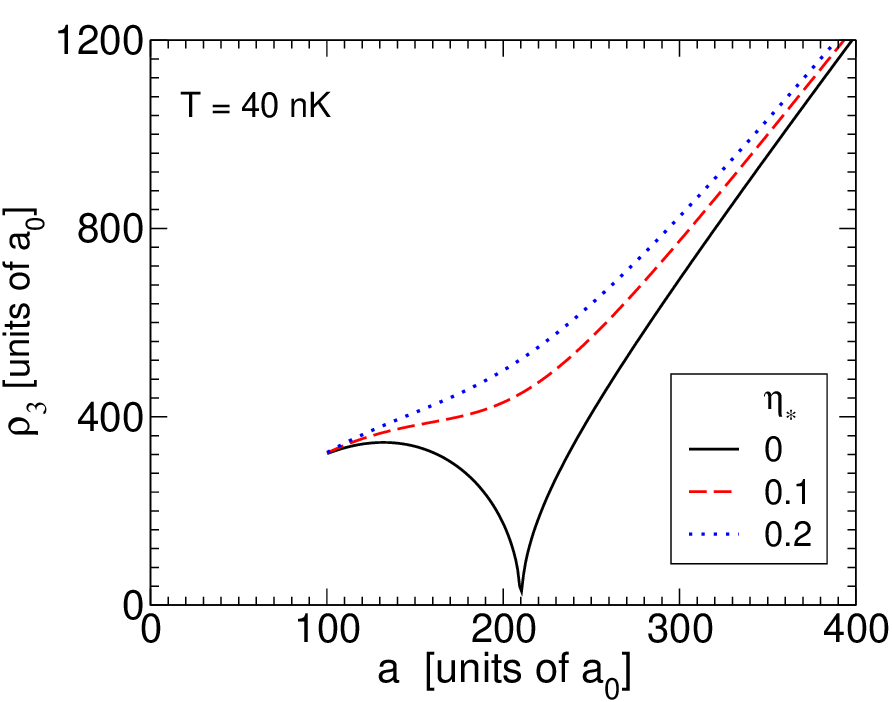}
\vspace*{0.0cm}
\caption{(Color online) The 3-body recombination length $\rho_3$ 
for $^{133}$Cs atoms as a function of $a$ for
$T=200$ nK (left panel) and $T=40$ nK (right panel).
The data points for $T=200$ nK are from Ref.~\cite{Grimm06}.
The curves are the zero-range predictions for three values 
of $\eta_*$: 
0 (solid lines), 0.1 (dashed lines), and 0.2 (dotted lines).}
\label{fig:alpha-CsT}
\end{figure}

In Fig.~\ref{fig:alpha-Cs}, we compare the zero-range predictions 
for the recombination length $\rho_3$ defined in Eq.~(\ref{rho3})
with the Innsbruck data for $T=200$~nK in the $a>0$ region.
A blow-up of the region $0 < a < 400~a_0$ is shown in 
the left panel of Fig.~\ref{fig:alpha-CsT}.
We assume $n_{\rm lost} = 3$, so that $L_3 = 3 \alpha$.
We set $a_{*0} = 210~a_0$, which is the central value of $a_{\rm min}$
measured in the Innsbruck experiment.  
In Fig.~\ref{fig:alpha-Cs}, we plot $\rho_3$ as a function of $a$ 
for $T = 200$ nK and three values of $\eta_*$: 0, 0.1 and 0.2.
The predictions for nonzero $\eta_*$ are larger than those for  
$\eta_*=0$ in the region $a < 500~a_0$ and smaller in the region $a > 500~a_0$.
We extend the zero-range predictions down to $a = 100~a_0$, 
which is smaller than the effective range of Cs atoms,
even though range corrections are expected to be 100\% 
at such small values of $a$.  The predictions for $\eta_*=0$ 
give a good fit to the data only in the region near $500~a_0$ 
where the predictions are insensitive to $\eta_*$.
The prediction for $\eta_*=0.2$ gives a remarkably good fit to the data 
for $a$ ranging from the highest point at $1228~a_0$ 
all the way down to $202.7~a_0$. As $a$ decreases below $200~a_0$, 
the prediction for $\eta_*=0.2$ continues to decrease monotonically 
while the data seems to have a local maximum near $137.3~a_0$.
The best fit to the data for $a$ in the range from  
$202.7~a_0$ to $1228~a_0$ is $\eta_* = 0.18$, 
which gives a $\chi^2$ per data point of 1.22.

%%%%%%%%%%%%%%%%%%%%%%%%%%%%%%%%%%%%%%%%%%%%%%%%%%
%%%%%%%%%%%%%%%%%%%%%%%%%%%%%%%%%%%%%%%%%%%%%%%%%%

In the right panel of Fig.~\ref{fig:alpha-CsT}, we show the
zero-range prediction for the lower temperature of $T=40$~nK.
For $\eta_*=0$, the local minimal of $\rho_3$
at $210~a_0$ becomes much deeper at 40~nK.
The predictions for $\eta_*=0.1$ and  $\eta_*=0.2$
at 40~nK are essentially indistinguishable from the
predictions at 200~nK in the region $a < 600~a_0$.

%%%%%%%%%%%%%%%%%%%%%%%%%%%%%%%%%%%%%%%%%%%%%%%%%%
\begin{figure}[htb]
\centerline{\includegraphics*[width=12cm,angle=0,clip=true]{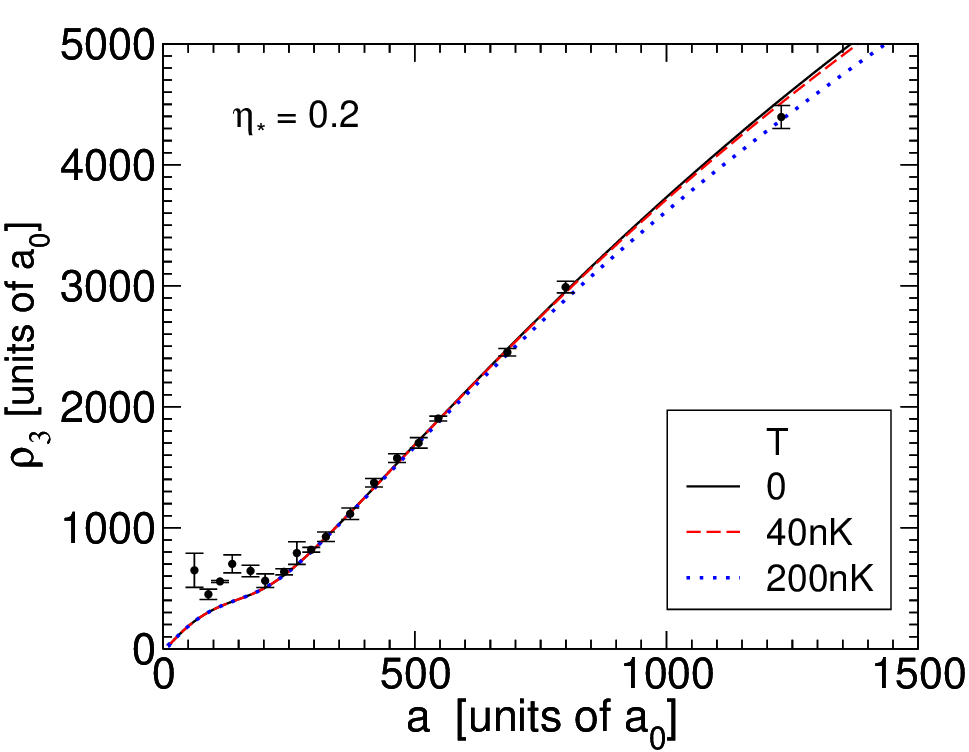}}
\vspace*{0.0cm}
\caption{(Color online) The 3-body recombination length $\rho_3$ 
for $^{133}$Cs atoms as a function of $a$ for $\eta_*= 0.2$.
The curves are the zero-range predictions for three values 
of the temperature: $T=0$ (solid line), 40~nK (dashed line),
and 200~nK (dotted line).
The data points are for $T=200$ nK from Ref.~\cite{Grimm06}.}
\label{fig:alpha-Cs:T}
\end{figure}
%%%%%%%%%%%%%%%%%%%%%%%%%%%%%%%%%%%%%%%%%%%%%%%%%%

In Fig.~\ref{fig:alpha-Cs:T}, we illustrate the temperature
dependence of the recombination length.
We show $\rho_3$ for $\eta_* = 0.2$ as a function of $a$ for
three temperatures: 0, 40~nK, and 200~nK.
Also shown is the Innsbruck data for 200~nK.
For $a < 500~a0$, the difference between the predictions at
$T=0$ and 200~nK is very small.  At larger values of $a$,
the prediction is significantly lower at 200~nK.
The prediction for $T=200$~nK fits the data much better than the 
prediction for $T=0$.
By comparing Figs.~\ref{fig:alpha-Cs} and \ref{fig:alpha-Cs:T},
we can see that increasing $\eta_*$ and increasing $T$ both 
have the effect of decreasing the prediction for $\rho_3$ 
for $a > 500~a_0$ without changing the prediction for smaller values of $a$.  
Thus accurate measurements of the temperature are essential 
if the data in this region is to be used to determine $\eta_*$.

\section{Comparison with previous work}

In this section, we compare our results with previous calculations 
of the 3-body recombination rate for identical bosons.
The previous results include calculations using specific models 
for the interactions between the atoms as well as calculations based 
on the universality of atoms with large scattering length.

Esry and D'Incao have calculated the 3-body recombination rate at nonzero
temperature by solving the 3-body Schr\"odinger equation for identical
bosons interacting through the simple potential
$V(r) = D/\cosh^2(r/r_0)$ \cite{DSE-04,DSE-07}.
In Ref.~\cite{DSE-07}, the depth parameter $D$ was used to vary the
scattering length from $+ \infty$ to $- \infty$ in the region with one
2-body bound state and from $+ \infty$ to $- \infty$ in the region with
two bound states.  This range of $D$ includes three universal regions
with large scattering length.
The range parameter $r_0$ determines the two 3-body
parameters $\kappa_*$ and $\eta_*$ for each of those universal regions.
To apply this simple model to the Innsbruck experiment on $^{133}$Cs
atoms, Esry and D'Incao tuned $r_0$ to fit the position of the
recombination resonance at $a = - 850~a_0$.
The predicted height of this resonance is
larger than that measured by the Innsbruck group.
Since $\eta_*$ is sensitive to the height and width of the resonance,
tuning $r_0$ to get the correct value of $\kappa_*$ gives too small
a value for $\eta_*$.  
Esry and D'Incao
also compared the predictions of their model to the Innsbruck data on
$^{133}$Cs atoms with a large positive scattering length. 
The model does not give a good quantitative fit to the 200~nK data for 
large positive scattering length. 

Massignan and Stoof have calculated the 3-body recombination rate at
nonzero temperature by solving the Skorniakov--Ter-Martirosian equation
in a scattering model for the lowest hyperfine spin state of
$^{133}$Cs atoms  \cite{MS07}.
Their scattering model is defined by a
2-body T-matrix element with 4 parameters
that models the Feshbach resonance of $^{133}$Cs atoms
centered at $B = -11$~G.  The 4 parameters consist of
the large background scattering length $a_{\rm bg} = 1800~a_0$ and
3 Feshbach resonance parameters: $B_0$, $\Delta B$, and $\Delta \mu$.
Their two-body T-matrix
also depends on a background effective range parameter $r_e$,
which provides a ``physical'' ultraviolet
cutoff and guarantees a well-behaved three-body integral equation.
Massignan and Stoof took $r_e$ to be a function of the Feshbach
resonance parameters: $r_e = -2\hbar^2/(m a_{\rm bg} \Delta \mu \Delta B)$.
The effective range is defined to be twice the coefficient of
$k^2$ in the low-momentum expansion of $k \cot \delta(k)$,
where $\delta(k)$ is the S-wave phase shift.
In their model, the effective range is always small
and negative, varying from $2r_e$ at the resonance to
$r_e \approx -0.3~a_0$ far from the resonance.
An estimate of the true background effective range
can be obtained by exploiting
the fact that the interatomic potential is $-C_6/r^6$ at long distances
and using $a_{\rm bg}$ and the van der Waals coefficient
$C_6$ for Cs atoms as inputs \cite{gao98,fgh99}:
$r_{\rm bg} \approx +250~a_0$.
Thus the model of Massignan and Stoof may not provide an accurate
representation of the 2-body problem.
For $T = 10$ nK, the prediction for the recombination length $\rho_3$
at the peak of the resonance in their model is too large by 40\%,
which is about 4 standard deviations. In the
nonuniversal region of small scattering length, the model predicts a
minimum in the 3-body recombination rate near $-150~a_0$ that was not
observed in the experiment. For positive scattering length,
the model does not give a good quantitative fit to the 200~nK data.
It predicts the local minimum to be at $240~a_0$, which is a little larger
than the observed position $210(10)~a_0$.  At the minimum, the prediction
for $\rho_3$ is less than half the measured value, which is too small
by more than 10 standard deviations.  At $a = 1228~a_0$,
their prediction for the recombination length $\rho_3$ is too large by
10\%, which is about 2 standard deviations.

Lee, K\"ohler, and Julienne have calculated the 3-body 
recombination rate at $T=0$ using a coupled-channel 
potential model for the relevant hyperfine spin states of $^{133}$Cs 
atoms \cite{LKJ07}. 
Their model had 6 fitting parameters: two interchannel
coupling parameters that were tuned to fit the position and width 
of the Feshbach resonance and 4 parameters in a separable interatomic 
potential that were tuned to fit the background scattering length,
the background effective range, and the binding energies of the 
two shallowest deep dimers \cite{LKJ07}.  They compared their 
zero-temperature calculations with the 
measurements of the Innsbruck group at nonzero temperature.
The agreement between their calculation and the Innsbruck data is
not very good even at a qualitative level. Their prediction for the
magnetic field of the recombination resonance in the $a<0$ region
is 5.1~G compared with the measured value of 7.4~G \cite{grimm1107}.
Away from the resonance, their predictions for $(\rho_3)^4$ 
are sometimes lower than the 10~nK data by about 2 orders of magnitude.
In the $a>0$ region, their predictions for $(\rho_3)^4$ 
were below the 200~nK data by at least one standard deviation
at all points and by many standard deviations at some of the points.
Their prediction for the position of the recombination minimum is 
about $130~a_0$, which is much smaller than the observed value
$210(10)~a_0$.

Lee, K\"ohler, and Julienne have criticized universal approaches to the
3-body recombination as being ``incomplete'' 
and they claimed that these approaches 
``could therefore lead to unreliable conclusions'' \cite{LKJ07}. 
These criticisms were based on
misconceptions. In the case of $^{133}$Cs atoms, the two deep dimers that
are closest to the threshold have an avoided crossing near the Feshbach
resonance at $-11$~G.  Lee, K\"ohler, and Julienne pointed out that as a
consequence the probabilities $Z_{-1}(B)$ and $Z_{-2}(B)$ for these deep
dimers to consist of atoms in the lowest hyperfine spin state change
dramatically with the magnetic field. They concluded incorrectly that
$K_{\rm deep}(0)$ must depend on $B$ not only through the
scattering length $a(B)$ but also through the strong $B$-dependence of the
properties of the deep dimers. It should be noted that no quantitative
evidence for this conclusion was given in Ref.~\cite{LKJ07}. Their conclusion
would certainly be true of the exclusive recombination rates into each of
the deep dimers. What the authors of Ref.~\cite{LKJ07} 
did not appreciate is that the
inclusive recombination rate summed over all deep dimers is much less
sensitive to details involving short distances, such as level crossings
between the deep dimers. The inclusive effects of the deep dimers can be
taken into account through the 3-body parameter $\eta_*$. 
Universal approaches are certainly ``incomplete'' in
the sense that, although they can describe the inclusive recombination rate
into deep dimers and the exclusive recombination rate
into the shallow dimer in the case $a>0$, they cannot describe
the exclusive recombination rates into individual deep dimers. 
However, contrary to the suggestion of Ref.~\cite{LKJ07}, 
there is no reason to expect universal approaches to lead to 
``unreliable conclusions'' for those recombination rates
that they can describe.
 
In Ref.~\cite{LKJ07}, Lee, K\"ohler, and Julienne 
referred to universal approaches to the 3-body problem  
as ``universal fitting procedures''. 
Assuming that the scattering length is known, the  zero-range results
require fitting the 3-body parameters $\kappa_*$ and $\eta_*$. 
This approximation is very economical 
in the number of fitting parameters. One price
that is paid for this economy is that the 3-body parameters depend
sensitively on the details of the 2-body potential. They may be so
sensitive that the only practical way to determine them is through
experimental measurements of 3-body observables.
The model used in Ref.~\cite{LKJ07} has 6 parameters that were 
used to fit results from a more accurate representation 
of the 2-body problem \cite{chin04}. 
Three of the parameters are sufficient 
to reproduce the scattering length $a(B)$ as a function of the 
magnetic field.  The other 3 were used to fit the background 
effective range and the binding energies of the two shallowest 
deep dimers.  The authors expected this to be sufficient to
calculate the 3-body recombination rate accurately.
The large errors in their predictions for the positions 
of the recombination resonance for $a<0$ and the recombination 
minimum for $a>0$ shows that this is not the case.
It implies that their model does not correctly predict the 
physical values of $\kappa_*$ for these two universal regions. 
One way to proceed would be to add more and more parameters 
to the interatomic potential and use them to fit additional 2-body
observables.  However  $\kappa_*$ and $\eta_*$ might be too 
sensitive to the details of interactions at short distances 
to make this approach practical.  An alternative that is 
motivated by the universal approach is to add
2 parameters to the interatomic potential and use them to tune
appropriate 3-body observables in the universal region 
to their measured values, thus tuning $\kappa_*$ and $\eta_*$
close to their physical values.

In Refs.~\cite{Braaten:2006qx,Shepard:2007gj,Platter:2007sn},
the 3-body recombination rate for the case $a>0$ has been 
calculated using simplifying assumptions to neglect some of the 
universal scaling functions.
The starting point in Ref.~\cite{Braaten:2006qx} was an approximation 
in which only the first term in the expression for $K^{(0)}(E)$ 
in Eq.~(\ref{Kshallow-S}), which corresponds to the 3-body channel 
labelled by $n=3$, was retained.
Inserting the zero-range expression for the S-matrix element 
$S_{AD,AAA}^{(J=0,3)}(E)$ in Eq.~(\ref{RL:ADAAA}),
this approximation becomes
%----------------------
\begin{equation}
K^{(0)}(E) \approx \frac{144 \sqrt{3} \pi^2}{x^4}
\left| s_{23}(x) 
+ \frac{s_{21}(x) s_{13}(x) e^{2i \theta_{*0}}}
      {1 - s_{11}(x) e^{2i \theta_{*0}}} \right|^2
\frac{\hbar a^4}{m}.
\label{Kshallow-trunc}
\end{equation}
%----------------------
%[e,hw,d,l]
This approximation could have been avoided by using 
Eq.~(\ref{sigbr:rl}) for $K^{(0)}(E)$ as a starting point instead.
The first simplifying assumption of Ref.~\cite{Braaten:2006qx}
was to neglect $s_{11}(x)$ in the denominator.
According to the analytic result in Eq.~(\ref{s11:x->0}),
this function is very small at the threshold:
$|s_{11}(0)| \approx 0.002$.  
The assumption of Ref.~\cite{Braaten:2006qx}
was that $|s_{11}(x)|$ remains small for larger values of $x$.
This assumption has now been verified by our explicit calculation 
of $s_{11}(x)$, which shows that  $|s_{11}(x)| < 0.035$ for $x < 10$.
With this simplifying assumption, the expression for the 
recombination rate in Eq.~(\ref{Kshallow-trunc}) can be reduced to
%----------------------
\begin{equation}
K^{(0)}(E) \approx \frac{144 \sqrt{3} \pi^2}{x^4}
\left| s_{23}(x)  e^{-i \theta_{*0}}
+ s_{21}(x) s_{13}(x) e^{i \theta_{*0}} \right|^2
\frac{\hbar a^4}{m}.
\label{Kshallow-simp}
\end{equation}
%----------------------
%[e,hw,d,l]
The amplitude inside the absolute value sign can be expressed 
as a linear combination of $\cos \theta_{*0}$ and $\sin \theta_{*0}$
with coefficients that are complex functions of $x$.
The coefficient of $\sin \theta_{*0}$ can be made real-valued by
multiplying the amplitude by an $x$-dependent phase that does not 
affect the rate.
Thus the approximation in Eq.~(\ref{Kshallow-simp}) depends on 
3 independent real-valued scaling functions.
The second simplifying assumption of Ref.~\cite{Braaten:2006qx}
was that the coefficients of $\cos \theta_{*0}$ and 
$\sin \theta_{*0}$ were relatively real.
This assumption was motivated by the 
existence of a deep minimum  in $K^{(0)}(E)$ at $E = 2.1$~mK
in the 3-body recombination rate for $^4$He atoms 
calculated in Ref.~\cite{SEGB02}.
This behavior can be most easily understood if the 
square of the complex amplitude in Eq.~(\ref{Kshallow-simp})
can be approximated by the square of a real 
amplitude that vanishes at $E = 2.1$~mK.
Given the second simplifying assumption, the approximation 
in Eq.~(\ref{Kshallow-simp}) depends on only two 
independent scaling functions. In Ref.~\cite{Braaten:2006qx}, 
these two scaling functions were 
constrained by the requirement that they reproduce the 
result of Ref.~\cite{SEGB02} for  $K^{(0)}(E)$ for 
$^4$He atoms interacting through the HFD-B3-FCI1 potential, 
which corresponds to $a = 1.15~a_{*0}$. By assuming
the absence of a near cancellation between the two scaling 
functions for $a = 1.15~a_{*0}$, 
they obtained predictions for the recombination rate with 
surprisingly narrow error bands in the region of $a$ near $a_{*0}$.
In Ref.~\cite{Shepard:2007gj}, Shepard reduced the error bands to a line
by using $K^{(0)}(E)$ for $^4$He atoms interacting through a second 
potential to determine the two scaling functions separately.
In Ref.~\cite{Platter:2007sn}, Platter and Shepard
removed the second simplifying assumption of Ref.~\cite{Braaten:2006qx}   
by using $K^{(0)}(E)$ for four different $^4$He potentials to determine 
the three independent real-valued scaling functions
in Eq.~(\ref{Kshallow-simp}).
Their approximation to the zero-range result for $K^{(0)}(E)$
is accurate to better than 5\%.

In Refs.~\cite{Braaten:2006qx} and \cite{Platter:2007sn}, 
the universal scaling functions that were extracted from 
calculations of 3-body recombination for $^4$He atoms 
were applied to $^{133}$Cs atoms with $a>0$.
The effects of deep dimers on the 3-body recombination rate 
into the shallow dimer were obtained by making the 
substitution in Eq.~(\ref{logsub}) for $\log a_{*0}$ 
in the expression for $K^{(0)}(E)$ in Eq.~(\ref{Kshallow-simp}).
Since their approach was unable to determine the 
function $s_{12}(x)$ that appears in Eq.~(\ref{Kdeep-s}),
they approximated the 3-body recombination rate 
$K_{\rm deep}(E)$ into the deep dimer
by its value at $E=0$, which is given in Eq.~(\ref{K-deep:approx}).
In Ref.~\cite{Braaten:2006qx}, the error bands 
associated with undetermined universal scaling functions
were small for $a < 300~a_0$ but they widened rapidly as $a$ increased,
so they were unable to obtain a quantitative fit to the Innsbruck data.
In Ref.~\cite{Platter:2007sn}, the authors obtained a reasonably 
good fit to the Innsbruck data for $a>200~a_0$ with values of $\eta_*$
in the range from 0 to 0.01, although their results tended to lie 
above the highest data point at $1228~a_0$.
Our definitive zero-range results give a significantly better fit. 
There is a qualitative difference in the dependence on $\eta_*$ 
between our definitive zero-range results and the
approximation in Ref.~\cite{Platter:2007sn}.
As $\eta_*$ increases, 
the definitive zero-range prediction in the region $a>600~a_0$ decreases 
as shown in Fig.~\ref{fig:alpha-Cs},
while the approximation in Ref.~\cite{Platter:2007sn} increases.
The difference comes primarily from the 
approximation of $K_{\rm deep}(E)$ by its threshold value
in Ref.~\cite{Platter:2007sn}.
As illustrated in Fig.~\ref{fig:alphadeep}, this is a good 
approximation for $E < 0.1~E_D$, but $K_{\rm deep}(E)$ decreases
rapidly with $E$ for energies above $E_D$.
The calculation of the energy dependence of $K_{\rm deep}(E)$
was essential for obtaining the excellent fit 
to the recombination length of $^{133}$Cs atoms with $\eta_* = 0.2$ 
that is illustrated in Fig.~\ref{fig:alpha-Cs}
and the left panel of Fig.~\ref{fig:alpha-CsT}.

\section{Summary and Outlook}

In this work, we have calculated the universal scaling functions
that determine the 3-body recombination rate for identical bosons
as a function of the collision energy in the
zero-range limit\footnote{Numerical results
for the universal scaling functions can be obtained by contacting
any of the authors.}. We used the STM equation
to calculate the phase shifts for elastic atom-dimer scattering
above the breakup threshold for angular momenta up to $J=6$.
For $J\ge 1$, the contribution $K^{(J)}(E)$ to the 3-body 
recombination rate into the shallow dimer is expressed in terms 
of a real-valued universal scaling function $f_J(x)$ in 
Eq.~(\ref{alpha-J:uni}). We determined $f_J(x)$ 
for each $J\ge 1$ by fitting the imaginary part of the 
atom-dimer phase shift.
For $J=0$, the contribution $K^{(0)}(E)$ to the 3-body 
recombination rate into the shallow dimer is expressed in terms 
of three complex-valued universal scaling functions 
$s_{11}(x)$, $s_{12}(x)$, and $s_{22}(x)$ in 
Eq.~(\ref{sigbr:rl}). We determined these scaling functions
by fitting the S-wave atom-dimer phase shift to Efimov's radial law. 
The parameters in the zero-range predictions for the 3-body
recombination are the scattering length and the 3-body parameter $a_{*0}$.
We compared our results to previously published calculations 
of the 3-body recombination rate of $^4$He atoms 
for angular momenta up to $J=3$ \cite{SEGB02}.
We used the 3-body recombination rate at threshold calculated 
in Ref.~\cite{SEGB02} to determine $a_{*0}$ for $^4$He atoms.  
We found good quantitative agreement for the 3-body recombination rate 
$K^{(J)}(E)$ as a function of energy in the $J=0$, 1, and 2 channels.

Our results can also be applied to atoms with large scattering length 
that have deep dimers.  The deep dimers provide additional 3-body 
recombination channels and they also have an effect on the 3-body 
recombination into the shallow dimer.  
An additional 3-body parameter $\eta_*$ is necessary to take into 
account the inclusive effects of the deep dimers. 
Remarkably, the 3-body recombination rate is completely determined 
by the same universal scaling functions that apply at $\eta_*=0$.
The $J=0$ contribution $K^{(0)}(E)$ to the 3-body 
recombination rate into the shallow dimer is expressed in terms 
of the three complex-valued universal scaling functions 
$s_{11}(x)$, $s_{12}(x)$, and $s_{22}(x)$ in Eq.~(\ref{Kshallow-s}).
The $J\ge 1$ contributions $K^{(J)}(E)$ are unaffected by the deep 
dimers and are still given by Eq.~(\ref{alpha-J:uni}). 
The inclusive rate $K_{\rm deep}(E)$ for 3-body 
recombination rate into deep dimers is expressed in terms 
of the universal scaling functions 
$s_{11}(x)$ and $s_{12}(x)$ in Eq.~(\ref{Kdeep-s}).
We compared our results to measurements by
the Innsbruck group of the 3-body recombination rate for
$^{133}$Cs atoms with a large positive scattering length 
at a temperature of 200~nK.
We set the 3-body parameter $a_{*0}$ equal to the position of the 
recombination minimum measured by the Innsbruck group \cite{grimm1107}:  
$a_{*0} = 210~a_0$. 
We then used the 3-body parameter $\eta_*$ to fit the dependence 
of the recombination length on $a$.
We find that $\eta_* = 0.2$ gives excellent agreement with the 
experimental results for $T=200$~nK in the region $a > 200~a_0$.
We also found that lowering the temperature would have an observable 
effect only for $a > 500~a_0$.

Our excellent fit to the $T=200$~nK data is particular remarkable
considering that it was obtained by using only the scattering length
and two adjustable 3-body parameters.
The accuracy of this fit presents a challenge to approaches
that use more fitting parameters but tune them to fit 2-body observables
in addition to the scattering length, as in Ref.~\cite{LKJ07}.
The 3-body parameters $\kappa_*$ and $\eta_*$ associated with a given
universal region are complicated functions of the 2-body parameters.
The 3-body parameters $\kappa_*$ and $\eta_*$ may be too sensitive 
to details at short distances to be determined
accurately by fitting a small number of 2-body  observables.
It is possible that the only practical way to determine them accurately
is by tuning them to fit measurements of 3-body observables.

Our zero-range results for the 3-body recombination rate have a
fractional theoretical uncertainty $r_s/|a|$,
where $r_s$ is the effective range.
Higher accuracy could be obtained by calculating range corrections.
It is known how to account systematically for these
corrections using an effective field theory for short-range interactions
\cite{Hammer:2001gh,Platter:2006ev}.  A first calculation of the 
range corrections to the recombination rate into the shallow dimer 
at threshold and an application to $^4$He atoms has already been carried
out \cite{Hammer:2006zs}. 
In the future, the range corrections should be calculated for
the recombination rate as a function of the energy.
The range corrections to the recombination rate into deep dimers
should also be calculated.
The range corrections are particularly important 
near a recombination minimum. In particular, 
they can shift the position of the minimum \cite{Hammer:2006zs}. 
In the case of Cs atoms, the importance of range corrections 
near the recombination minimum is further
increased by the fact that the minimum is at a scattering length
$210~a_0$ that is not large compared to the effective range.

Our calculations and those in Refs.~\cite{DSE-07,MS07}
have been carried out in the Boltzmann region of temperature and density.
It would be useful to have zero-range predictions for lower
temperatures where the Bose-Einstein distribution has to be used.
In the Boltzmann region, we only need the hyperangular average 
$K(E)$ of the 3-body recombination rate, which can be determined from the
atom-dimer phase shifts using unitarity.  In the Bose-Einstein region,
it is necessary to calculate the 3-body recombination rate directly.
This requires the calculation of scattering rates
with 3 atoms in the initial state.  
The relevant S-matrix elements are expressed in terms of
universal scaling functions $s_{11}(x)$, $s_{12}(x)$, $s_{1n}(x)$, 
and $s_{2n}(x)$ in Eq.~(\ref{RL:ADAAA}).
The scaling functions $s_{1n}(x)$ and $s_{2n}(x)$ would have to be
calculated for those 3-atom states labelled by $n=3,4,\ldots$ 
that are most important near the 3-atom threshold.

It is also important to calculate the definitive zero-range predictions 
for the 3-body recombination rate for negative scattering length.
In this case, recombination can occur only into deep dimers. 
Previous calculations based on the universal approach
have used the adiabatic hyperspherical approximation
\cite{Jonsell06} or a resonance approximation \cite{YFT06}.
The relevant universal scaling functions can be determined by
calculating 3-atom elastic scattering rates near the 3-atom threshold.

\begin{acknowledgments}
This research was supported in part by the Department of Energy 
under grants DE-FG02-05ER15715 (EB) and DE-FG02-93ER40756 (LP),
by the National Science Foundation under Grant No.~PHY--0653312 (LP),
the UNEDF SciDAC Collaboration under DOE Grant DE-FC02-07ER41457 (LP),
by an Ohio University postdoctoral fellowship (LP),
by the DFG through funds provided to SFB/TR 16 (HWH),
by the BMBF under grant number 06BN411 (HWH),
and by the Korea Research Foundation under grant 
KRF-2006-612-C00003 (DK).

\end{acknowledgments}

\end{document}